\documentclass[10pt,journal,compsoc]{IEEEtran}
\usepackage{subfiles}
\usepackage{url}
\usepackage[pdftex]{graphicx}
\usepackage{amsmath}
\usepackage{xspace}
\usepackage{multirow}
\usepackage{amssymb}
\usepackage{algorithm}
\usepackage{subfig}
\usepackage[noend]{algpseudocode}
\renewcommand{\algorithmicrequire}{\textbf{Input:~}}
\renewcommand{\algorithmicensure}{\textbf{Output:~}}

\newcommand\myeq{\mkern1.5mu{=}\mkern1.5mu}
\usepackage{booktabs}
\usepackage{array}
\usepackage{color}
\usepackage{listings}
\usepackage{placeins}
\usepackage{enumitem}
\usepackage{balance}

\ifCLASSOPTIONcompsoc
  \usepackage[nocompress]{cite}
\else
  \usepackage{cite}
\fi

\ifCLASSINFOpdf
\else
\fi

\newcommand\acro{{\sc{HyperFuseR\xspace}\xspace}\xspace}
\usepackage{xcolor}

\begin{document}

\title{Fast and Error-Adaptive Influence Maximization based on Count-Distinct Sketches}

\author{G\"{o}khan~G\"{o}kt\"{u}rk
        and~Kamer~Kaya
\IEEEcompsocitemizethanks{\IEEEcompsocthanksitem G. G\"{o}kt\"{u}rk and K. Kaya are with Computer Science and Engineering, Faculty of Engineering and Natural Sciences, Sabanci University, TR~34956, Istanbul, Turkey.}
}


\IEEEtitleabstractindextext{%
\begin{abstract}
Influence maximization~(IM) is the problem of finding a seed vertex set that maximizes the expected number of vertices influenced under a given diffusion model. Due to the NP-Hardness of finding an optimal seed set, approximation algorithms are frequently used for IM. 
In this work, we describe a fast, error-adaptive approach that leverages Count-Distinct sketches and hash-based fused sampling. To estimate the number of influenced vertices throughout a diffusion, we use per-vertex Flajolet-Martin sketches where each sketch corresponds to a sampled subgraph. To efficiently simulate the diffusions, the reach-set cardinalities of a single vertex are stored in memory in a consecutive fashion. This allows the proposed algorithm to estimate the number of influenced vertices in a single step for simulations at once. 
For a faster IM kernel, we rebuild the sketches in parallel only after observing estimation errors above a given threshold. Our experimental results show that the proposed algorithm yields high-quality seed sets while being up to $119\times$ faster than a state-of-the-art approximation algorithm. In addition, it is up to $62\times$ faster than a sketch-based approach while producing seed sets with $3\%$--$12\%$ better influence scores.
\end{abstract}
} 

%
%
%
%
%
%
%
%
%

\maketitle

\IEEEdisplaynontitleabstractindextext
\IEEEpeerreviewmaketitle

\ifCLASSOPTIONcompsoc
\IEEEraisesectionheading{\section{Introduction}\label{sec:introduction}}
\else
\section{Introduction}
\label{sec:introduction}
\fi

Efficient information/influence dissemination in a network is an important research area with several applications in various fields, such as viral marketing~\cite{leskovec2007dynamics, trusov2009effects}, social media analysis~\cite{zeng2010social, moreno2004dynamics}, and recommendation systems~\cite{lu2012recommender}.
As the study of these networks is imperative for educational, political, economic, and social purposes, a high-quality seed set to initiate the diffusion may have vital importance.
Furthermore, since the diffusion analysis may be time-critical, or increasing the influence coverage may be too expensive, novel and efficient approaches to find good vertex sets that propagate the information effectively are essential.

Influence maximization is the problem of finding a subset $S \subset V$ of $K$ vertices in a graph $G = (V, E)$ with the vertex set $V$ and edge set $E$ such that $S$ reaches the maximum reachability, i.e., influences the maximum expected number of vertices, under some diffusion model. Kempe et al.~\cite{kempe2003maximizing} introduced the IM problem, proved it to be NP-hard, and provided a greedy Monte-Carlo approach that has a constant approximation ratio over the optimal solution. This greedy approach is one of the most frequently applied algorithms for IM. The time complexity of the greedy algorithm, with an influence score estimate $\sigma$, running $R$ simulations, and selecting $K$ seed vertices is $\mathcal{O}(KRn\sigma)$ for a graph with $n$ vertices. Although they perform well in terms of seed-set quality, the greedy Monte-Carlo solutions are impractical for real-life networks featuring millions of vertices as a consequence of their expensive simulation costs. Due to this reason, many heuristics and proxy methods have been proposed in the literature~\cite{MixGreedy, narayanam2010shapley, kimura2007extracting, chen2010PMIA,chen2010LDAG, kim2013scalable, cohen2014sketch, goyal2011simpath, jung2012irie,cheng2014imrank,liu2014influence,galhotra2016holistic}.

 

Simulating a greedy algorithm in parallel is a straightforward workaround to reduce the execution time of IM kernels and make them scalable for large-scale networks. However, for large networks, a parallel, greedy approach with a good approximation guarantee does not come cheap on networks with billions of vertices and edges even if a large number of processing units/cores are available. Following similar attempts in the literature, we propose a parallel, sketch-based approach that approximates the Monte-Carlo processes. To boost the performance, the proposed approach does not exactly count the number of influenced vertices. Instead, it leverages Count-Distinct sketches. Below is a summary of our contributions:

\begin{itemize} [leftmargin=0.3cm]
\item We propose \acro\footnote{\scriptsize{\url{https://github.com/ggokturk/infuser}}}, an open-source, blazing-fast, sketch-based and accurate Influence Maximization algorithm. The proposed scheme samples the edges as they are traversed across several simulations. Thus, sampling, diffusion, and count-distinct processes are fused for all simulations. 

\item Running concurrent simulations on per-vertex Count-Distinct sketches reduces the number of memory accesses which is the main bottleneck for many graph kernels in the literature. While traversing an edge, \acro concurrently performs multiple diffusion simulations while using only a single (8-bit) value per vertex for each simulation. 

\item \acro{} can process large-scale graphs with millions of vertices and hundreds of millions of edges under a minute without compromising the quality of results. Furthermore, the performance scales near linearly with the number of threads available. In addition, while processing a large-scale graph, only a few GBs of memory is used, where most of the memory is spent for storing the graph itself. 

\item Once it is read from the memory, \acro processes all samples of a single edge together. The suggested approach, therefore, decreases the pressure on the memory subsystem. Furthermore, it employs vector compute units to its near maximum efficiency to regularize memory accesses.

\item We evaluate the runtime performance, memory consumption, and influence score of sketch- and approximation-based state-of-the-art influence maximization algorithms, namely {\sc Skim}~\cite{cohen2014sketch}, {\sc Tim+}~\cite{tim} and {\sc Imm}~\cite{minutoli2019fast}, to accurately position the performance of \acro{} within the IM literature. The experiments show that \acro can be $62\times$ and $119\times$ faster than a state-of-the-art sketch-based and approximation algorithm, respectively while reaching the influence quality of the accurate algorithms with less memory.

\end{itemize}

The paper is organized as follows: 
In Section~\ref{sec:background}, we present 
the background on IM and introduce the mathematical notation. 
Section~\ref{sec:method} describes the proposed approach in detail.
In Section~\ref{sec:evaluation}, a thorough performance evaluation is provided by conducting experiments on various real-world datasets and influence settings. A detailed empirical comparison with the state-of-the-art from the literature is also given. Section~\ref{sec:relatedwork} presents a comparative overview of the existing work. Finally, Section~\ref{sec:conclusion} discusses future work and concludes the paper.

\section{Notation and Background}\label{sec:background}

Let $G = (V,E)$ be a directed graph where the $n$ vertices in $V$ represent the agents, and $m$ edges in $E$ represent the relations among them. An edge $(u,v) \in E$ is an {\em incoming} edge for $v$ and an {\em outgoing} edge of $u$. The {\em incoming} neighborhood of a vertex $v \in V$ is denoted as $\Gamma^-_{G}(v) = \{u: (u,v) \in E\}$. Similarly, the {\em outgoing} neighborhood of a vertex $v \in V$ is denoted as $\Gamma^+_{G}(v) = \{u: (v,u) \in E\}$. A graph $G' = (V',E')$ is a sub-graph of $G$ if $V' \subseteq V$ and $E' \subseteq E$. The diffusion probability on the edge $(u, v) \in G$ is noted as $w_{u,v}$, where $w_{u,v}$ can be determined either by the diffusion model or according to the strength of $u$ and $v$'s relationship.

\begin{table}[!ht]
    \caption{Table of notations}
    \label{tab:notation}
    \centering
    \begin{tabular}{|l|p{0.7\linewidth}|}
        \hline
        Variable & Definition  \\
        \hline
        $G = (V,E)$     & Graph $G$ with vertices $V$ and edges $E$ \\
         $\Gamma^+_G(u)$ & The set of vertices $v$ where $(u,v) \in E$ \\ 
        $\Gamma^-_G(u)$  &The set of vertices $v$ where $(v, u) \in E$\\ 
        $w_{u,v}$       & Probability of $u$ directly influencing $v$ \\
        $R_{G}(v)$      & Reachability set of vertex $v$ on graph $G$\\
        \hline\hline
        $S$             & Seed set to maximize influence\\
        $K$             & Size of the seed set\\
        ${\cal J}$   & Number of Monte-Carlo simulations performed\\
        $\sigma_{G}(S)$ & The influence score of $S$ in $G$, i.e., expected number of vertices reached from $S$ in $G$\\
        \hline\hline
        $w_{u,v}$             & Sampling probability for the edge $(u,v)$\\
        $P(s,v)_r $     & Random probability generated for selecting edge vertices $s$ to $v$ in simulation $r$\\
        $h(u,v)$        & Hash function for edge $\{u,v\}$\\
        $h_{max}$       & Maximum value hash function $h$ can return\\
        \hline\hline
        $e$             & Estimated reachability set size\\
        $M_u[j]$        & $j$th sketch register for vertex $u$\\
        $\varsigma $    & Influence gained before last sketch build\\
        $\sigma $       & Influence Score\\
        $\delta$        & Marginal gain after last sketch build\\
        $err_l$         & Local estimation error of the sketch\\
        $err_g$         & Global estimation error of the sketch\\
        $\epsilon_{g}$    & Global estimation error threshold\\
        $\epsilon_{l}$    & Local estimation error threshold\\ 
        $\epsilon_{c}$    & Non-convergenced vertex threshold\\
        \hline         
    \end{tabular}
\end{table}
\subsection{Influence Maximization}

Influence Maximization aims to find a seed set $S \subseteq V$ among all possible size $K$ subsets of $V$ that maximizes an {\em influence spread function} $\sigma$  when the diffusion process is initiated from $S$. 
In the literature, {\em independent} and {\em weighted cascade}~(IC and WC), and 
{\em linear threshold}~(LT)~\cite{kempe2003maximizing} are three widely recognized diffusion models for IM. 

\begin{figure}[!ht] 
    \centering
  \subfloat[\small{IC}\label{fig:ic}]{%
       \includegraphics[width=0.37\linewidth]{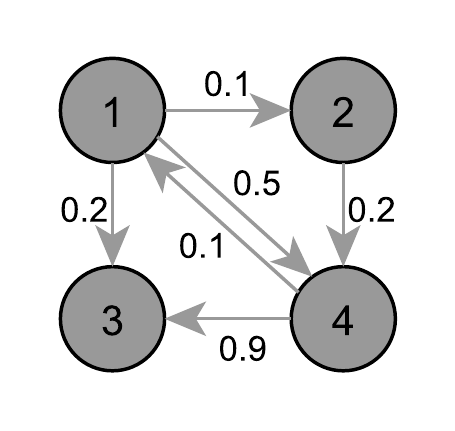}}
  \subfloat[\small{WC}\label{fig:wc}]{%
        \includegraphics[width=0.37\linewidth]{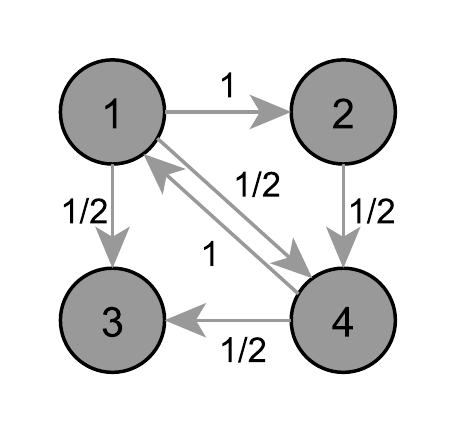}}
    \\
  \caption{\small{\protect\subref{fig:ic} 
The directed graph $G = (V, E)$ for IC with independent diffusion probabilities. 
\protect\subref{fig:wc}
The directed graph for WC is obtained by setting the diffusion probabilities of incoming edges to $1 / |\Gamma^-_G(v)|$ for each vertex $v \in V$.
  }}
  \label{fig:xx} 
\end{figure}

\begin{itemize}[leftmargin=*]
\item The {\bf Independent Cascade} model works in rounds and activates a vertex $v$ in the current round if one of $v$'s incoming edges $(u, v)$ is used during the diffusion round, which happens with the activation probability $w_{u, v}$, given that $u$ has already been influenced in the previous rounds. The activation probabilities are independent (from each other and previous activations) in the {\em independent cascade} model, which we focus on in this paper. A toy graph with activation probabilities on the edges is shown in Figure~\ref{fig:ic}.
In theory, there can exist parallel and independent $\{u, v\}$ edges in $E$. In practice, they are merged to a single $\{u,v\}$ edge via preprocessing. 

\item The {\bf Weighted Cascade}  model is a variant of the independent cascade that uses the structural properties of vertices to set the edge weights as shown in Figure~\ref{fig:wc}.
The method, as described in~\cite{kempe2003maximizing}, sets $w_{u, v} = 1 / d_v$ where $d_v$ is the number 
of incoming edges of $v$~(which in the original graph is equal to $\Gamma^-_G(v)$).
Therefore, if $v$ has $\ell$ neighbors activated in the last round, its probability of activation in the new round is $1-( 1-1 / d_v)^\ell$. 


\item{\bf Linear threshold} generalizes the IC model and activates the vertex $v$ once the cumulative activation coming from its neighbors exceeds a given threshold $\theta_v$. 
All the $(u, v)$ edges with active $u$ vertices are taken into account in the process. Vertex $v$ is activated when the total activation probability through these edges exceeds $\theta_v$~\cite{kempe2003maximizing}.  
\end{itemize}

The complexity analysis stays consistent for many diffusion models, including {\em Independent Cascade}, {\em Weighted Cascade}, and {\em Linear Threshold} models; the time complexity of the greedy algorithm, estimating the $\sigma$ influence score, running $R$ simulations, and selecting $K$ seed vertices is $\mathcal{O}(KRn\sigma)$ for a graph with $n$ vertices. 
We concentrate on the IC model in this paper, but although their adaptation requires some work, the proposed methods are also relevant to other models in the literature.

\subsection{Count-Distinct Sketches}\label{sec:sketch}
The {\em distinct element count} problem focuses on finding the number of distinct elements in a stream where the elements are coming from a universal set ${\cal U}$. Finding the number of vertices to be influenced of a candidate seed vertex $u$, i.e., the cardinality of $u$'s {\em reachability set}, is a similar problem. For each sample subgraph, the number of visited vertices is found while traversing the subgraphs starting from $u$. Note that an exact computation of set cardinality requires memory proportional to the cardinality, which is ${\mathcal O}(|{\cal U}|)$. 

The reachability set of a vertex is the union of all its connected vertices (via outgoing edges). Many IM kernels exploit this property to some degree. The methods based on {\em reverse reachability} \cite{borgs2014maximizing} 
utilize this property directly to merge the reachability sets of connected vertices to estimate the number of vertices influenced. {\em MixGreedy} \cite{MixGreedy} 
goes one step further; it utilizes the fact that for an undirected graph, all vertices in a connected component have the same reachability set. Therefore, all the reachability sets within a single sample subgraph can be found via a single graph traversal. 

For directed graphs, storing reachability sets for all vertices and merging these sets are infeasible for nontrivial graphs. 
If one-hot vectors are used to store the reachability sets for constant insertion time, $\mathcal{O}(n^2\mathcal{R})$ bits of memory is required where each merge operation has $\mathcal{O}(n)$ time complexity. 
If disjoints sets are used for storing reachability sets; $\mathcal{O}(n{\sigma}\mathcal{R})$ 
memory is required to store all reachability sets, and each merge operation has $\mathcal{O}({\tt Ack}(\sigma))$ complexity where ${\tt Ack}$ is the Ackermann~\cite{ackermann} function. 

Count-Distinct Sketches can be leveraged to estimate reachability sets' cardinality efficiently; for instance, the Flajolet–Martin~(FM) sketch~\cite{flajolet1985probabilistic} can do this with a constant number, ${\cal J}$, of registers. Furthermore, the union of two sketches can be computed in constant time. The FM sketch stores that how rare the elements are in a stream. The rarity of the elements is estimated by counting the maximum number of leading zeros in the stream elements' hash values. Initially, each register is initialized with zero. The items are hashed one by one, and the length of the longest all-zero prefix is stored in the register. With a single register, the cardinality estimation can be done by computing the power $2^\ell$ where $\ell$ is the value in the register.

In practice, multiple registers and hash values, $M[j]$ and $h_j$, are commonly used to reduce variance. For a sketch with multiple registers, the impact of adding an item $x \in \cal{U}$ is shown in~\eqref{eq:sketch-add}:
\begin{equation}
    \label{eq:sketch-add}
    M[j] = \max(M[j], clz(h_j(x)),~1 \leq j \leq {\cal J}
\end{equation} where $clz(y)$ returns the number of leading zeros in $y$ and ${\cal J}$ is the number of sketch registers. With multiple registers, the average of the register values can be used to estimate the cardinality, and the result is divided to a correction factor $\phi \approx 0.77351$ to fix the error due to hash collisions. That is the estimated cardinality $e$ is computed as
\begin{equation}
    \label{eq:sketch-estimate}
    e = 2^{\bar{M}}/\phi
\end{equation} 
where $\bar{M} = {\tt{avg}}_j\{M[j]\}$ is the mean of the register values.

In this work, we utilize a variant of Flajolet–Martin sketch; since multiple Monte-Carlo simulations are performed to calculate the estimated influence, we use one register per simulation and take the average length of the longest leading zeros. Two given FM sketches $M_u$ and $M_v$ can be merged, i.e., their union $M_{uv}$ can be computed by taking the pairwise maximums of their registers. Formally; 
\begin{equation}
\label{eq:sketch-merge}
    M_{uv}[j] = \max(M_u[j], M_v[j]),~1 \leq j \leq {\cal J}.
\end{equation} 
In our implementation, the merge operations are performed if and only if there is a sampled edge between the vertices. 

\section{Error-Adaptive Influence Maximization}\label{sec:method}

Most IM algorithms have the same few steps to find the best seed vertex set; sampling, building the influence oracle, verifying the impact of new candidates, and removing the latent seed set's residual reachability set. Following the idea proposed in~\cite{infuser}, \acro fuses the sampling step with other steps to avoid reading the graph multiple times. 

\acro first performs a diffusion process; the process starts with per-vertex sketches that are initialized with the hash value of the corresponding vertex~(i.e., every vertex reaches to itself). Then, for all the (sampled) edges ($u, v$), i.e., the ones that contribute to the diffusion for this simulation, the sketch of the source vertex $u$ is merged into that of the target vertex $v$ until all sketch registers for all the vertices converge. The merge operation utilized in this process is slightly different from the conventional one and retrofitted to mimic the IC diffusion. 

Throughout the process, each register is used only for a single sample/simulation. 
For an edge ($u, v$) in simulation $j$ where $v$ is a live vertex, an update, i.e., a merge operation, on $u$'s register is performed on the corresponding register $M_u$[$j$], i.e., $M_u[j] = max(M_u[j],M_v[j])$.
That is, at each iteration, vertices (outgoing) neighbors' reachability sets are added to their sketches.
This recursive formulation of the influence iteratively relays the reachability information among the vertices, allowing us to estimate the marginal influence for all vertices very fast.

After estimating the reachability set cardinalities, \acro picks the vertex $v$ with the largest cardinality by evaluating the sketches. Then it finds the (actual) reachability set of the latent seed set, which is the union of the reachability sets of $v$ and the vertices in the seed set, by performing Monte-Carlo simulations. The vertices in this reachability set are removed from the live set $L$. 
Hence, in later iterations, these vertices will not contribute to the marginal gain. Finally, the algorithm checks if rebuilding is necessary for the sketches based on the difference between the sketch estimate and Monte-Carlo estimate. 

\subsection{Hash-based Fused Sampling}
\begin{figure}[!ht] 
    \centering
    \subfloat[\label{fig:sims}]{%
        \includegraphics[width=0.5\linewidth]{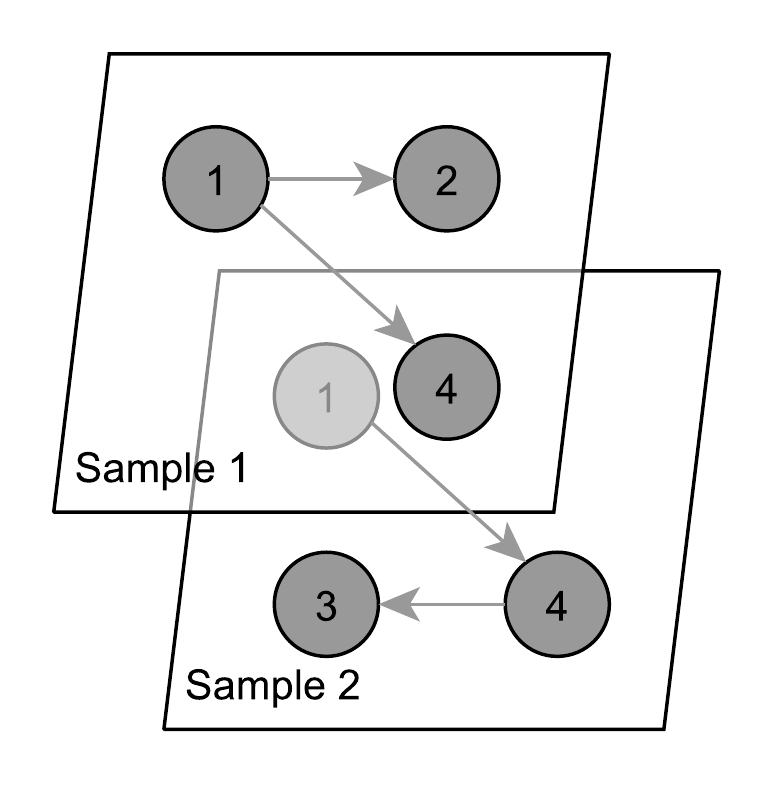}
    } 
    \subfloat[\label{fig:fused}]{%
     \includegraphics[width=0.5\linewidth]{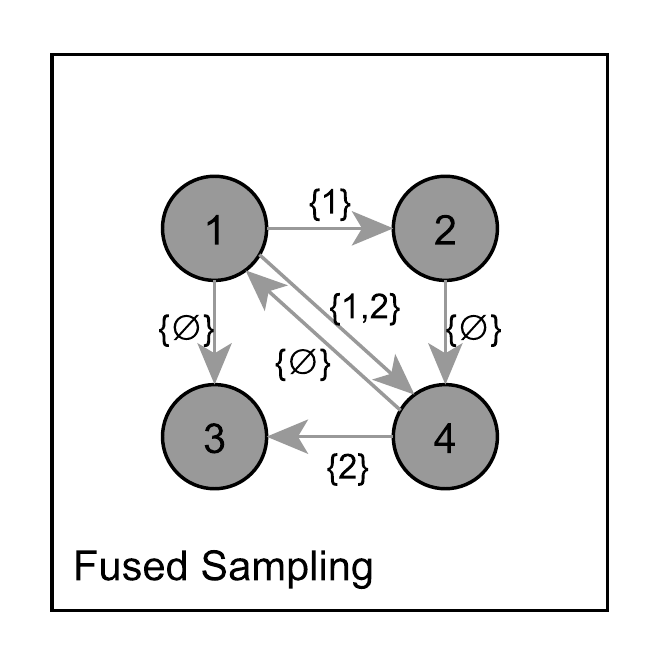}}
  \caption{\small{
  \protect\subref{fig:sims} Two sampled subgraphs of the toy graph from Figure~\ref{fig:ic} with 4 vertices and 6 edges.
  \protect\subref{fig:fused} The simulations performed are fused with sampling. Each edge is labeled with the corresponding sample/simulation IDs. 
  }}
  \label{fig:traversal} 
\end{figure}

The probabilistic nature of cascade models requires sampling subgraphs from $G = (V, E)$ to simulate the diffusion process. If performed individually as a preprocessing step, as the literature traditionally does, sampling can be an expensive stage, furthermore, a time-wise dominating one for the overall IM kernel. We identify two main bottlenecks; first, sampling multiple sub-graphs may demand multiple passes on the graph, which can be very large and expensive to stream to the computational cores, and second, if samples are memoized, the memory requirement can be a multiple of the graph size. 
In this work, we borrow the fused-sampling technique from {\sc InFuseR}~\cite{infuser} which eliminates the necessity of creation and storage of the sample subgraphs in memory. 
In Fig.~\ref{fig:traversal}, we briefly illustrate fused-sampling; instead of processing the samples independently as in Fig.~\ref{fig:sims}, fused-sampling processes each edge concurrently for multiple simulations as shown in Fig.~\ref{fig:fused}. This allows us to process each edge only a few times instead of once for every simulation.

In \acro, when an edge of the original graph is being processed, it is processed for all possible samples. Then, it is decided to be {\it sampled} or {\it skipped} depending on the outcome of the hash-based random value for each sample. Given a graph $G = (V, E)$, for an edge $(u, v) \in E$, the hash function used is given below:
\begin{equation}
    \label{eq:hash}
    h(u,v) = \mbox{{\sc Murmur3}}(u||v)~mod~2^{31}  
\end{equation}
where $||$ is the concatenation operator. In our preliminary experiments, we have tried a set of hash algorithms. After a careful analysis, we chose {{\sc Murmur3}} due to its simplicity and good avalanche behavior with a maximum bias $0.5\%$~\cite{MurmurHash3Performance}. Although the approach mentioned above generates a unique hash value for each edge, and hence a unique sampling probability, different simulations require different probabilities. First, a set of uniformly randomly chosen numbers $X_r \in_R [0, h_{max}]$ associated with each simulation $r$ are generated to enable this for each edge. Then the sampling probability of $(u, v)$ for simulation $r$, $P(u, v)_r$, is computed: To do this, the hash value, $h(u,v)$, is XOR'ed with  $X_r$ and the result is normalized by dividing the value to the upper limit of the hash value $h_{max}$. Formally,
\begin{equation}
    \label{eq:hash_prob}
    P(u,v)_r = \frac{X_r \oplus h(u,v)}{h_{max}}.
\end{equation}
The edge $(u,v)$ exists in the sample $r$ if and only if  ${P}(u,v)_r$ is smaller than the edge threshold $w_{u,v}$. One of this approach's benefits is that an edge can be sampled using a single XOR and compare-greater-than operation. Moreover, the corresponding control flow branch overhead can be removed using {\em conditional move} instructions.

\begin{figure}[!ht] 
    \centering
    \includegraphics[width=1\linewidth]{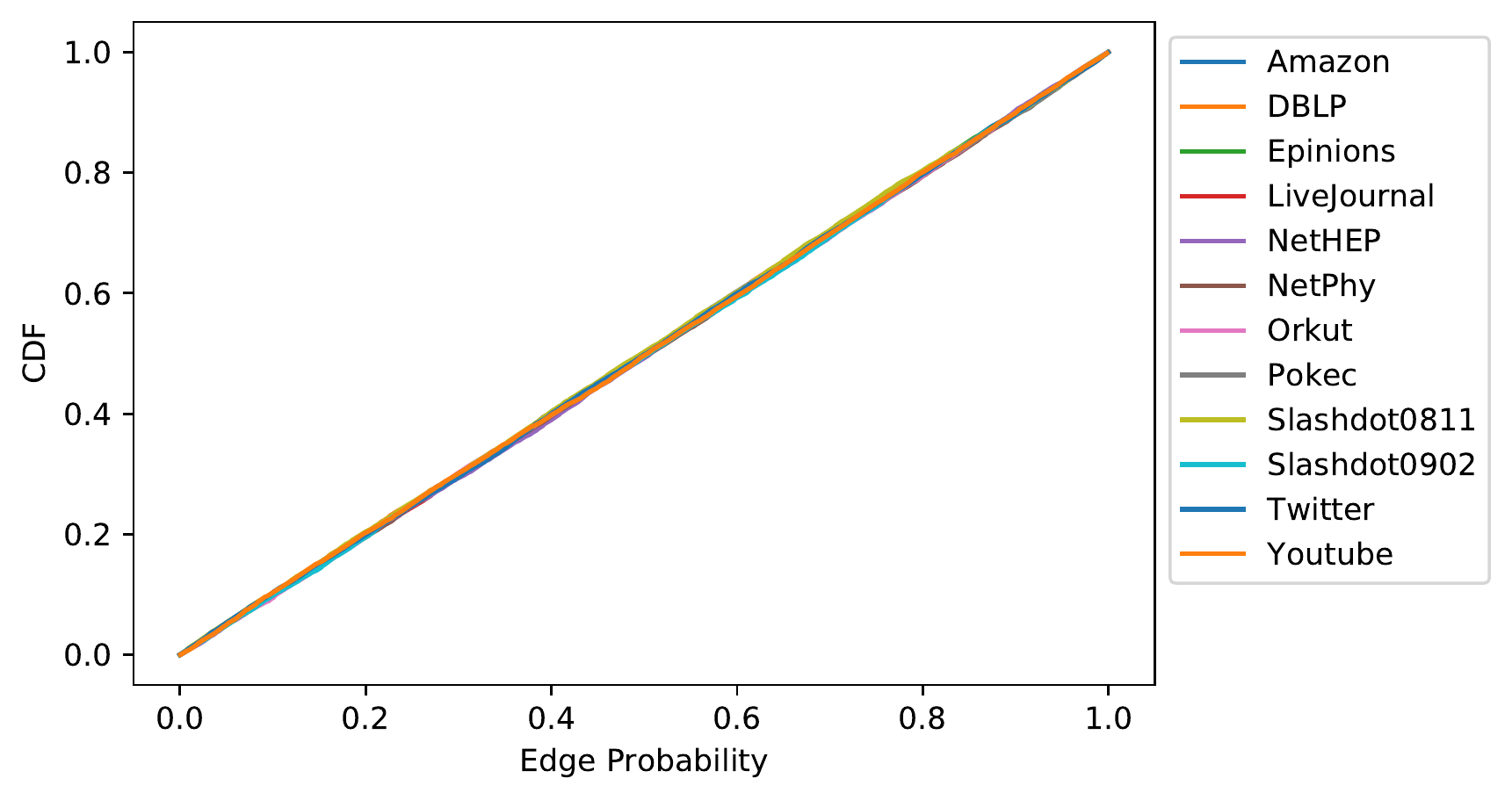}
    \caption{\small{Cumulative probability function of hash-based sampling probabilities on various real-life networks.}}
    \label{fig:prob-cdf} 
\end{figure}

\begin{figure}[!ht] 
    \centering
    \includegraphics[width=1\linewidth]{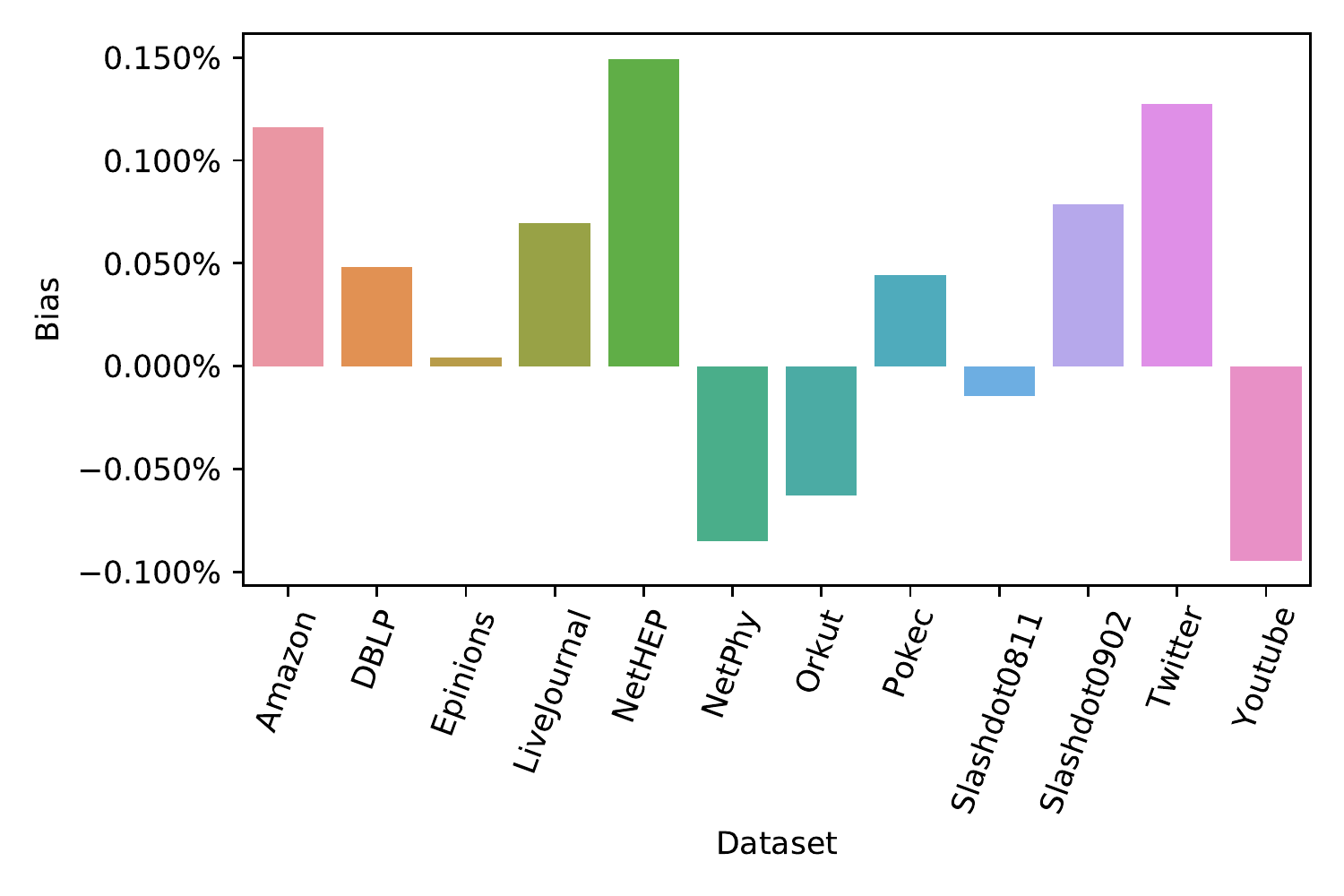}
    \caption{\small{Bias distribution of hash-based sampling probabilities on various real-life networks.}}
    \label{fig:prob-bias} 
\end{figure}

Using a strong hash function such as {{\sc Murmur3}} ensures that all bits independently change if the input is changed. This property allows us to generate good enough pseudo-random values for fair sampling. To evaluate the randomness and fairness of the values generated with the hash-based approach, we generated a large number of samples for various real-life networks and plotted the cumulative distribution~(Fig.~\ref{fig:prob-cdf}) and the bias of the random values $P(u,v)_r$ used~(Fig.~\ref{fig:prob-bias}). As the former shows, the sampling distribution of the hash-based computation values resembles a uniform random distribution. Furthermore, the latter shows that the bias is insignificant for each network. 

Being able to generate the samples on the fly allows us to avoid many memory accesses. The only downside of hash-based fused sampling is that we have to generate all these random values,   $P(u,v)_r$, for each edge traversal and each simulation $r$. The edges' hash values are precomputed for all the edges in $E$ to reduce computation cost and leverage fused sampling's performance gains. Fortunately, the rest of the operations, i.e., one XOR and one division, are very fast on modern computing hardware. 

\subsection{Estimating Reachability Set Cardinality}
A greedy solution to the influence maximization problem requires finding a vertex that maximizes the marginal influence gain at each step until the seed set size reaches $K$. For an exact computation, one must find all candidate vertices' reachability sets within all the samples. Such a task involves many graph traversals and is expensive even with various algorithmic optimizations and a scalable parallelized implementation, e.g., see~\cite{infuser}. The influence estimation problem is quite similar to the Count-Distinct problem applied to all sample subgraphs, as explained above. Hence, in this work, we pursue the idea of using Count-Distinct sketches to estimate marginal influence scores. In this work, we propose an efficient and effective IM kernel, \acro, that utilizes Flajolet–Martin sketches described in Section~\ref{sec:sketch} to estimate the averages of distinct elements in the sampled subgraphs. Algorithm~\ref{algo:main} shows the steps taken by the kernel.

\renewcommand{\baselinestretch}{0.95}
\begin{algorithm}
\caption{\sc{\acro}($G,K,{\cal J}$)}
\label{algo:main}
\algorithmicrequire{$G = (V,E)$: the influence graph
\\\hspace*{6.6ex}{$K$: number of seed vertices
\\\hspace*{6.7ex}${\cal J}$: number of Monte-Carlo simulations}\\}
\algorithmicensure{$S$: a seed set that maximizes the influence on $G$
}
\begin{algorithmic}[1]
    \State {$S \leftarrow \{\emptyset\}$}
    \For{$ v\in V$} {\bf in parallel}
        \For{$ j \in \{1, \ldots, {\cal J}\}$}
            \State $M_v[j] \leftarrow clz(hash(v) \oplus hash(j))$ 
        \EndFor
    \EndFor
    \State $M \leftarrow ${\sc Simulate}$(G,M,{\cal J},\emptyset)$
    \State $M_{S'} \leftarrow zeros({\cal J})$
    \State $\varsigma \leftarrow 0$
    \For{$k=1\ldots K$} \label{line:for}
        \State $s \leftarrow \underset{v\in V}{\mathrm{argmax}}\{${\sc Estimate}$(${\sc Merge}$(M_{S'},M_v))$\}\label{line:estimate}
        
        \State $S \leftarrow S \cup \{s\}$     
        \State $e \leftarrow ${\sc Estimate}$(${\sc Merge}$(M_{S'},M_s))$\label{line:e}
        \State $R_G(S) \leftarrow$ reachability set of $S$ (for all simulations)
        \State $\sigma \leftarrow$ Monte-Carlo-based (actual) influence of $S$
        \State $\delta = \sigma - \varsigma$
        \State $err_l=|(e - \delta) / \delta|$
        \State $err_g=|(e-\delta) / \sigma|$
        \If{$ err_l < \epsilon_l \lor err_g < \epsilon_g$} 
            \State $M_{S'} \leftarrow$ {\sc Merge}$(M_{S'},M_s)$ \label{line:if}
        \Else 
            \For{$ v\in V$} {\bf in parallel}\label{line:else1}
                \For{$ j \in \{1, \ldots, {\cal J}\}$}
                    \State $M_v[j] \leftarrow clz(hash(v) \oplus hash(j))$ 
                \EndFor
            \EndFor
            \State $M \leftarrow ${\sc Simulate}$(G,M,{\cal J},R_{G}(S))$
            \State $M_{S'} \leftarrow zeros({\cal J}) $ 
            \State $\varsigma \leftarrow \sigma $ \label{line:else2} 
        \EndIf
    \EndFor
    \State \Return $S$
\end{algorithmic}
\end{algorithm}
\renewcommand{\baselinestretch}{1}

\begin{figure*}[!ht]
    \begin{center}
    \includegraphics[width=0.7\linewidth]{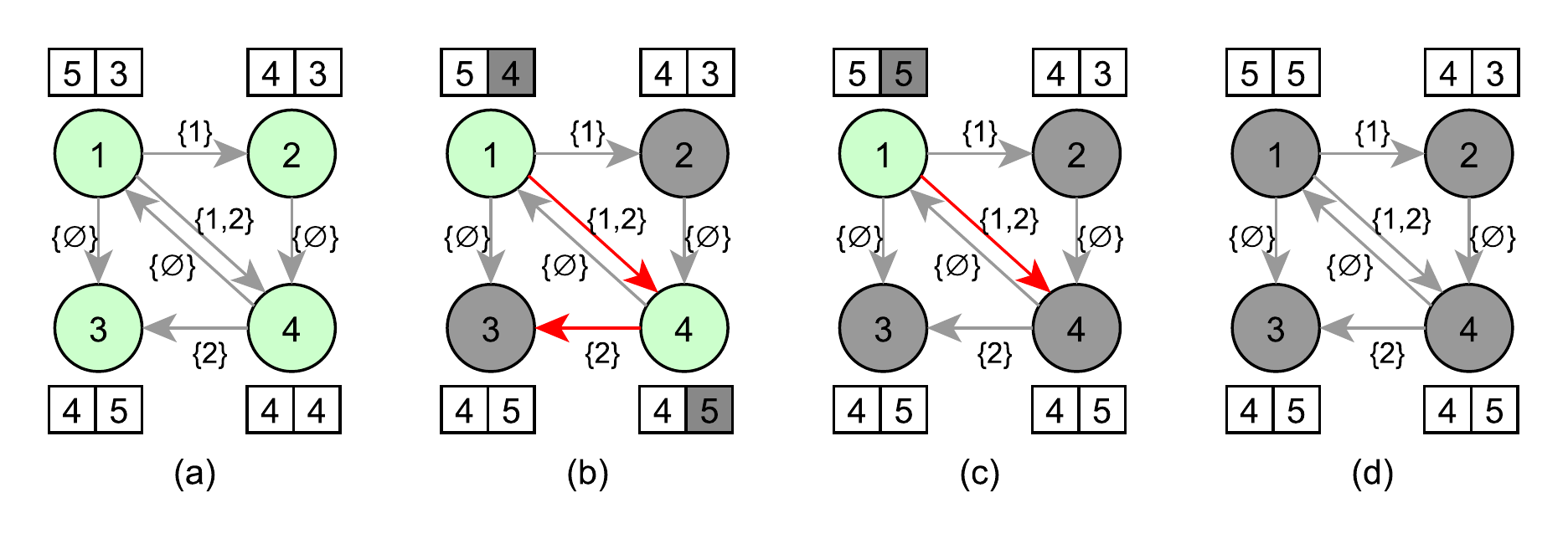}
    \caption{\small{(a) The initial state on the toy graph for \acro{}; all vertices are set as live~(green), and their registers are initialized with the length of the zero prefix of their hashes. (b) 
    For the ($1, 4$) edge which is live for both simulations, 1's registers are set to the maximum of both 1's and 4's registers. The ($4, 3$) edge is live only in the second simulation. Hence, the second register of $4$ is updated to $5$. For the second iteration, vertices 1 and 4 are live~(green) since their registers have changed. (c) For the live ($1, 4$) edge, 1's second is updated and 1 is set as live again. (d) All the registers converged. As no live vertices exist, the process stops.}}\label{fig:hf-processing} 
    \end{center}
    \end{figure*}

 Algorithm~\ref{algo:main} first initializes the reachability sets of all vertices by adding the vertices themselves. 
 That is for all vertices $u$, its $j$th register is set to $M_u[j]=clz(h_j(u))$ meaning $R_{{G}_j}(u) = \{u\}$ where $G_j$ is the $j$th sampled graph. 
 Then, we perform the diffusion process on the sketch registers whose pseudocode is given in Algorithm~\ref{algo:diffusion-step}. The diffusion starts by adding all the vertices to the {\em live vertex set} $L$. 
 Then at each step, the incoming edges of the live vertices are processed. 
 For a vertex $u$, its sketch, $M_u$, is updated by merging the sketches $M_v$ of all live outgoing neighbors vertices $v \in {L} \cap \Gamma^+_{G}(u)$. For each such vertex $v$ and simulation $j$, the operation $M_u[j] = max(M_u[j], M_v[j])$ is performed. This approach can be seen as a bottom-up, i.e., reversed,  diffusion process where at each iteration, the cardinality information is pulled from vertices neighbors.
If any of $u$'s sketch registers changes during this operation, it is added to the live vertex set $L'$ of the next iteration. Once the incoming edges of all live vertices are processed, the iteration ends. Figure~\ref{fig:hf-processing} shows how \acro performs two simulations at the same time using sketch registers.

\renewcommand{\baselinestretch}{0.95}
\begin{algorithm}[!ht]
\caption{\sc{Simulate}($G,M,{\cal J},R_S$)}
\label{algo:diffusion-step}
\algorithmicrequire{$G = (V,E)$: the influence graph
\\\hspace*{6.6ex}{$M$: sketch vectors of vertices
\\\hspace*{6.7ex}${{\cal J}}$: number of MC simulations
\\\hspace*{6.7ex}$R_S$: reachability set of the seed set
}
\\}
\algorithmicensure{$M$: updated Sketch vectors
}
\begin{algorithmic}[1]
    \State {$L \leftarrow V$}
    \State {$L' \leftarrow {\emptyset}$}
    \While{$|L|/|V| > \epsilon_c$}
        \For{$u \in \Gamma(L)$} {\bf in parallel} \label{ln:inner_start} 
        \For{$e_{u,v} \in A(u) $}
            \For{$j \in (0,{\cal J}]$} \label{ln:vec1}
                \If{$P(u,v)_j < w_{u,v} \land u  \not\in R_S[j]$} \label{ln:early_exit}
                    \State{$M_u[j]\leftarrow max(M_u[j],M_v[j])$}\label{ln:update}          
                \EndIf
            \EndFor
            \If {$M_u$ changed}
                \State $L' \leftarrow L' \cup u $ \label{ln:inner_end}
            \EndIf
        \EndFor
        \EndFor
        \State $L \leftarrow L'$
        \State $L' \leftarrow \{\emptyset\}$
    \EndWhile
    \State \Return $M$
\end{algorithmic}
\end{algorithm}
\renewcommand{\baselinestretch}{1}

The traditional Greedy algorithm~\cite{kempe2003maximizing} processes the simulations one-by-one and computes the vertices' reachability sets. On the other hand, \acro efficiently performs multiple simulations at once in a single-step iteration. Since each iteration relays one level of cardinality information, this step requires at most $d$ iterations where $d$ is the diameter of $G$. When processed individually as the Greedy algorithm does, the $j$th simulation over the sampled subgraph $G_j$ would require only at most $d_j$ iterations, which is the diameter of $G_j$, and probably much smaller than $d$. Although \acro seems to perform much more iterations, $d$ is a loose upper bound for \acro. A better one is {\tt max}$\{d_j: 1 \leq j \leq {\cal J}\}$ where $d_j$ is the diameter of $G_j$. To further reduce the overhead of concurrent simulations and avoid bottleneck simulations due to remaining perimeter vertices, we employ an early-exit threshold $\epsilon_c$ over the remaining live vertices ratio, which is expected to be very small when only one or two simulations remain. That is if $|L'| \leq |V| \times \epsilon_c$ the diffusion process in Algorithm~\ref{algo:diffusion-step} stops. Otherwise, $L$ is set to $L'$, $L'$ is cleared, and the next iteration starts. We used $\epsilon_c = 0.02$ to make \acro faster while keeping its quality almost the same.

After the diffusion process, the following steps are repeated until $K$ vertices are added to the seed set $S$. First, for each $v \in V$, the cardinality of the reachability set, $R_G(S \cup v)$, is estimated by merging $M_{S'}$ and $v$'s sketch registers where $M_{S'}$ is the set of sketch registers for the seed set $S$ used to estimate the number of already influenced vertices by $S$.\footnote{In fact, the definition is exact only if sketch rebuilding is disabled. As it will be described in the following subsection, when \acro's error-adaptive mechanism is enabled, $M_{S'}$ is periodically rebuilt to estimate the cardinality of reachability sets over the remaining, unblocked vertices. This is why $S'$ is used instead of $S$.} Before the kernel, these registers are initialized with zeros. Second, a vertex $s$ with the maximum cardinality estimation is selected and added to $S$. Third, the actual simulations are performed to compute the reachability set of $S$. Having an actual $R_G(S)$ allows us to calculate the estimation errors and find the blocked vertices for all simulations, which is vital since these blocked vertices can be skipped during the next diffusion steps. Besides, we leverage the actual influence to have an {\em error-adaptive kernel}, i.e., to compute the actual sketch error and rebuild the sketches when the accumulated error reaches a critical level which can deteriorate the quality for the following seed vertex selections.

\subsubsection{Error-adaptive sketch rebuilding}

Sketches are fast. However, each sketch operation, including update and merge, can decrease their estimation quality below a desired threshold. Our preliminary experiments revealed that sketches are highly competent at finding the first few seed vertices for influence maximization. Unfortunately, after a few seed vertices, the sketch registers $M_{S'}$, which are updated at line~\ref{line:if} of Algorithm~\ref{algo:main} via merging with new seed vertex $s$'s reachability set, become large. The saturation of $M_{S'}$ registers is important since \acro uses them to select the best seed candidate at line~\ref{line:estimate}. When they lose their sensitivity for seed selection, a significant drop in the quality is observed. 

\begin{figure}[!ht]
    \begin{center}
    \includegraphics[width=0.93\linewidth]{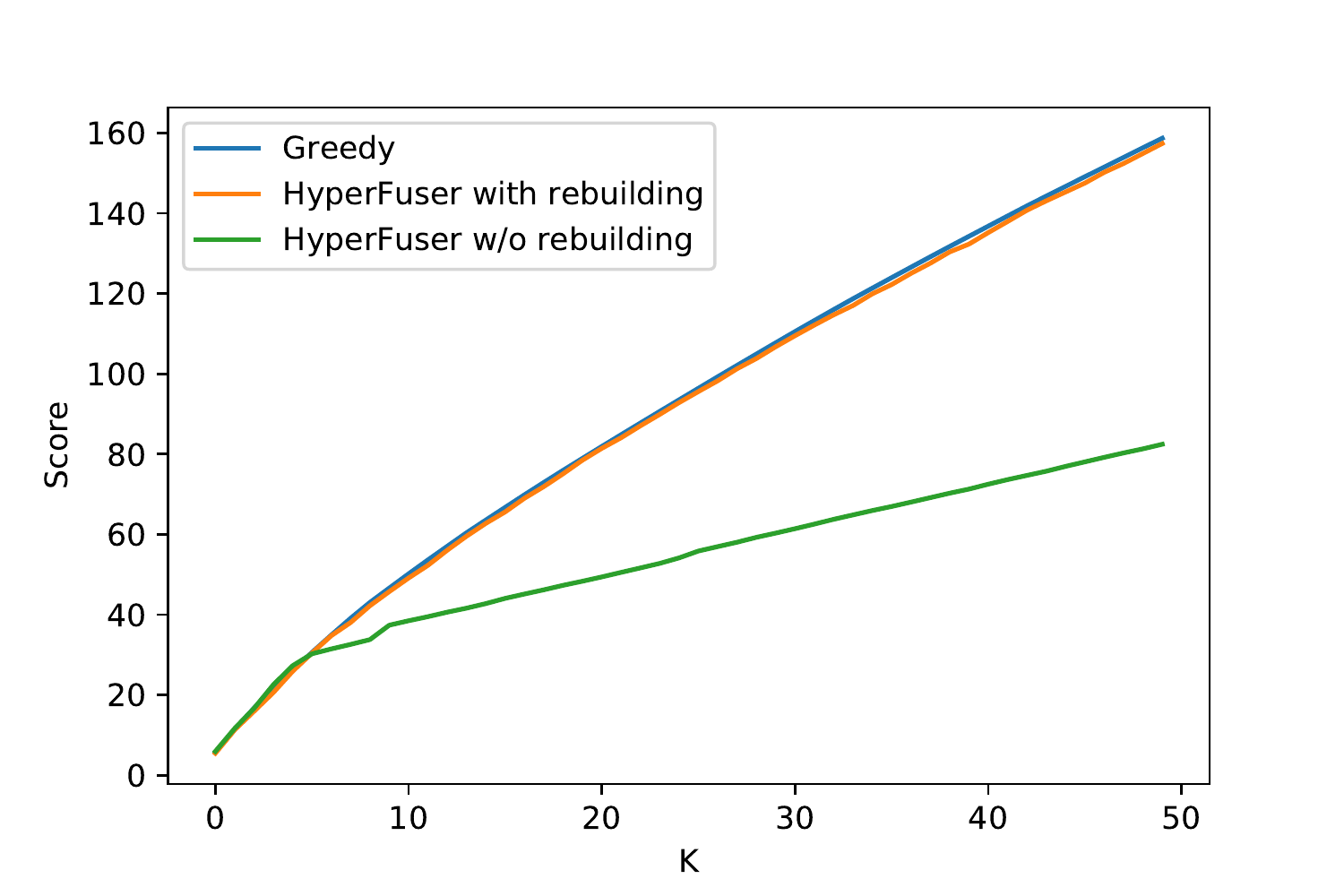}
    \caption{ Effect of register saturation on Amazon dataset using \acro(${\cal J}=256$) without rebuilding against Greedy($\mathcal{R}=20000$) method\cite{kempe2003maximizing}. 
     }\label{fig:sketch-saturation} 
    \end{center}
\end{figure}

\begin{figure*}[!ht]
    \begin{center}
        \subfloat[{\tt LiveJournal} execution time in seconds\label{fig:lj-time}]{%
        \includegraphics[width=0.33\linewidth]{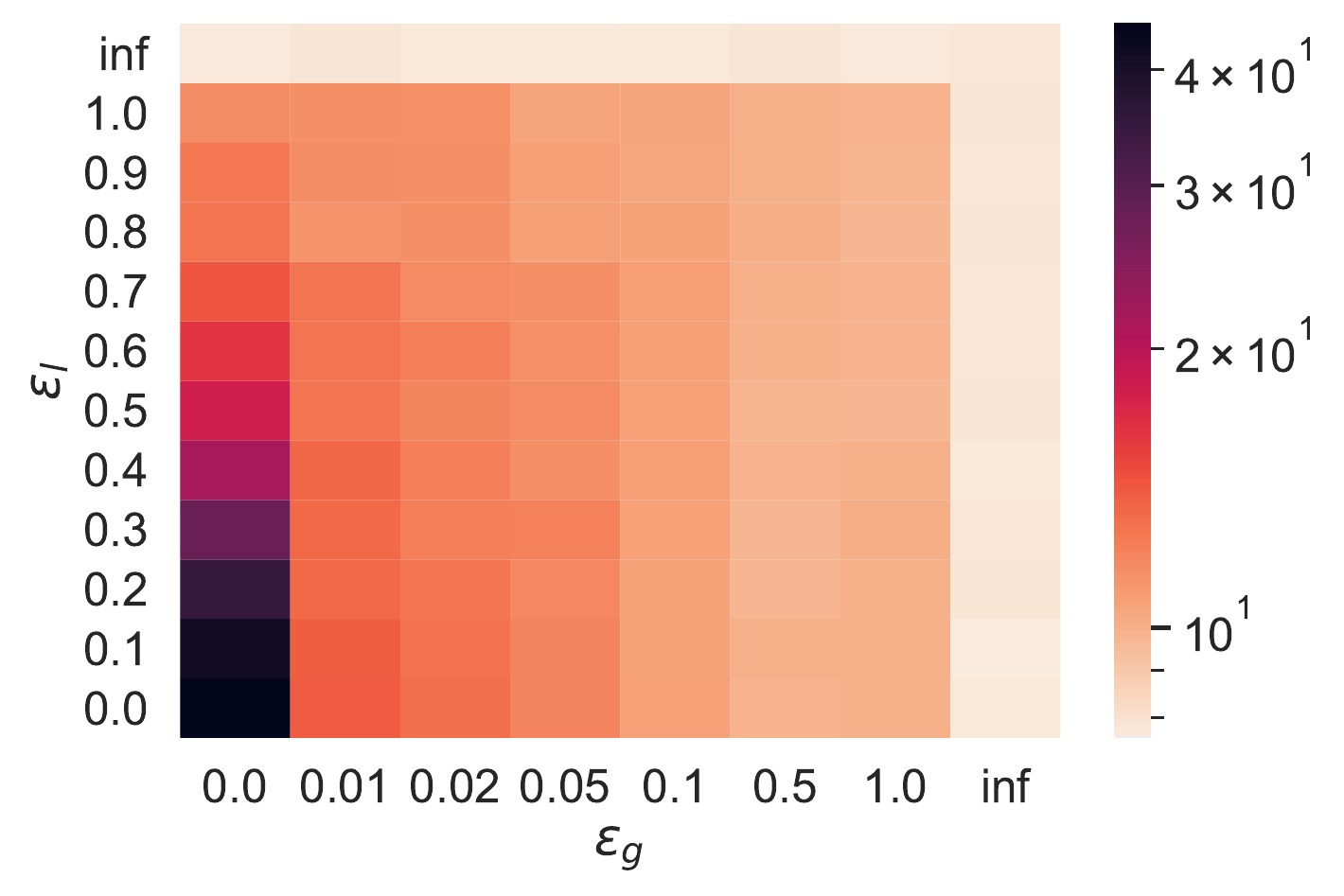}
    }
    \subfloat[{\tt Orkut} execution time in seconds\label{fig:orkut-time}]{%
        \includegraphics[width=0.33\linewidth]{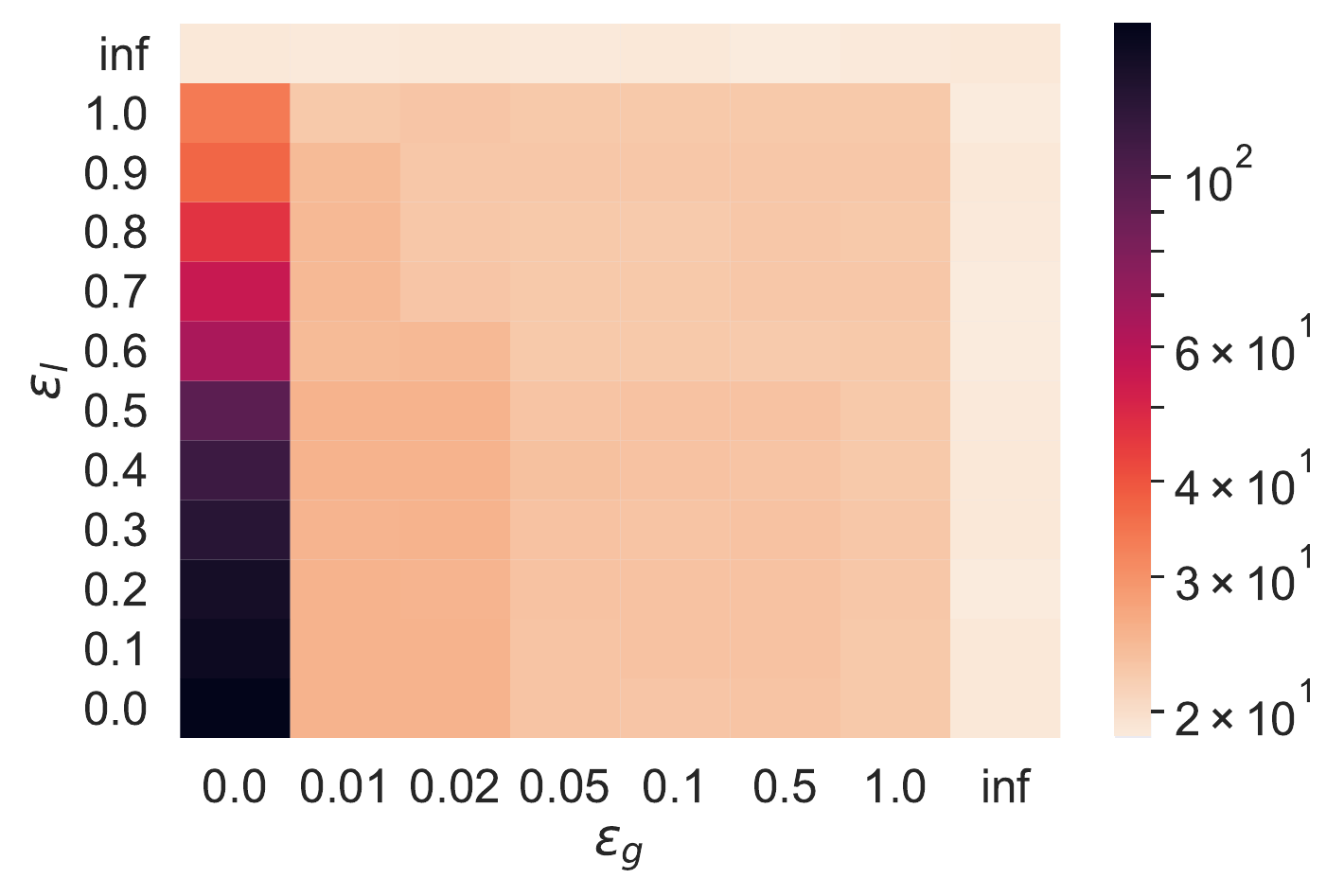}
        }
    \subfloat[{\tt Pokec} execution time in seconds\label{fig:pokec-time}]{%
        \includegraphics[width=0.33\linewidth]{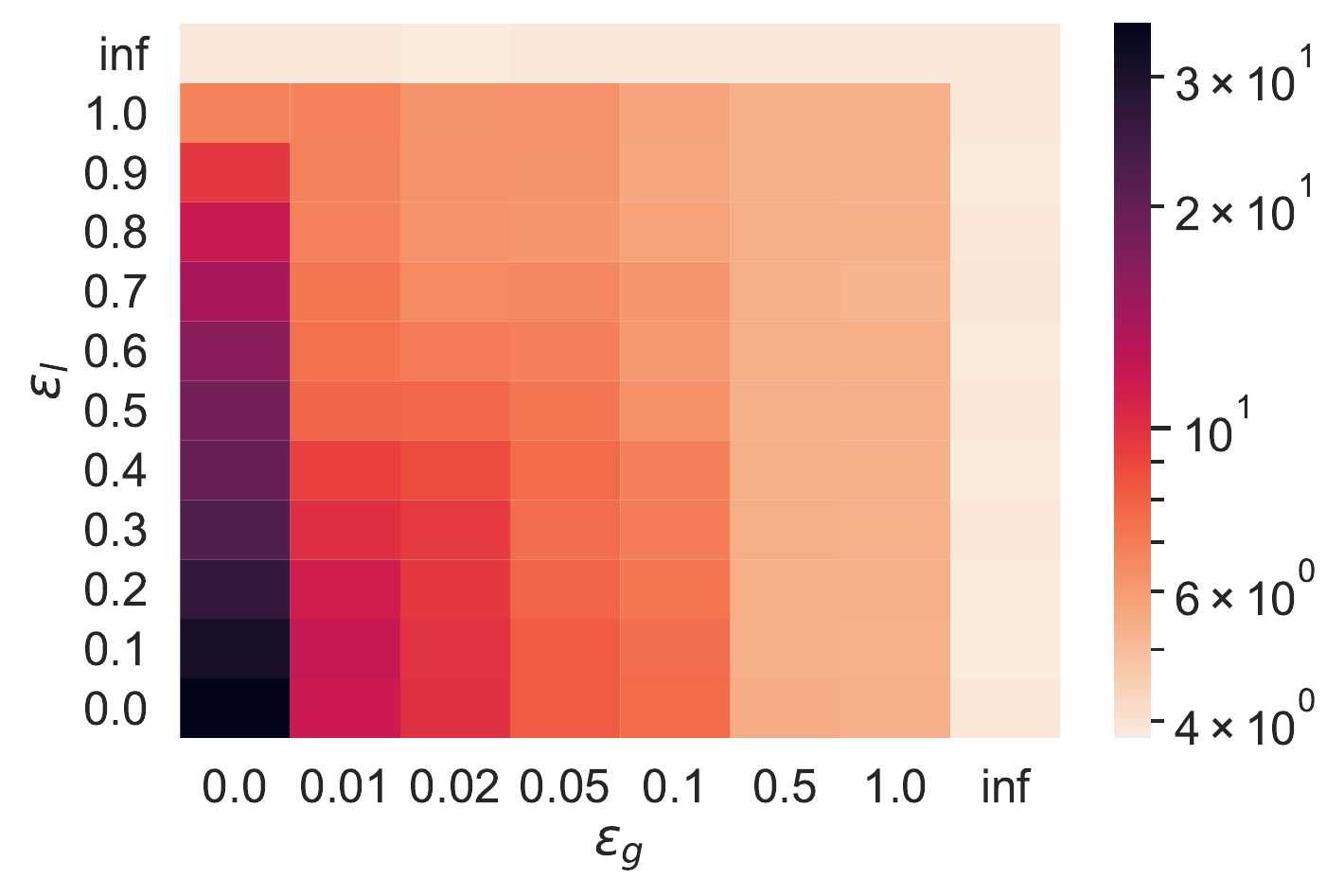}
        }\\
        \subfloat[{\tt LiveJournal} influence score\label{fig:lj-score}]{%
        \includegraphics[width=0.33\linewidth]{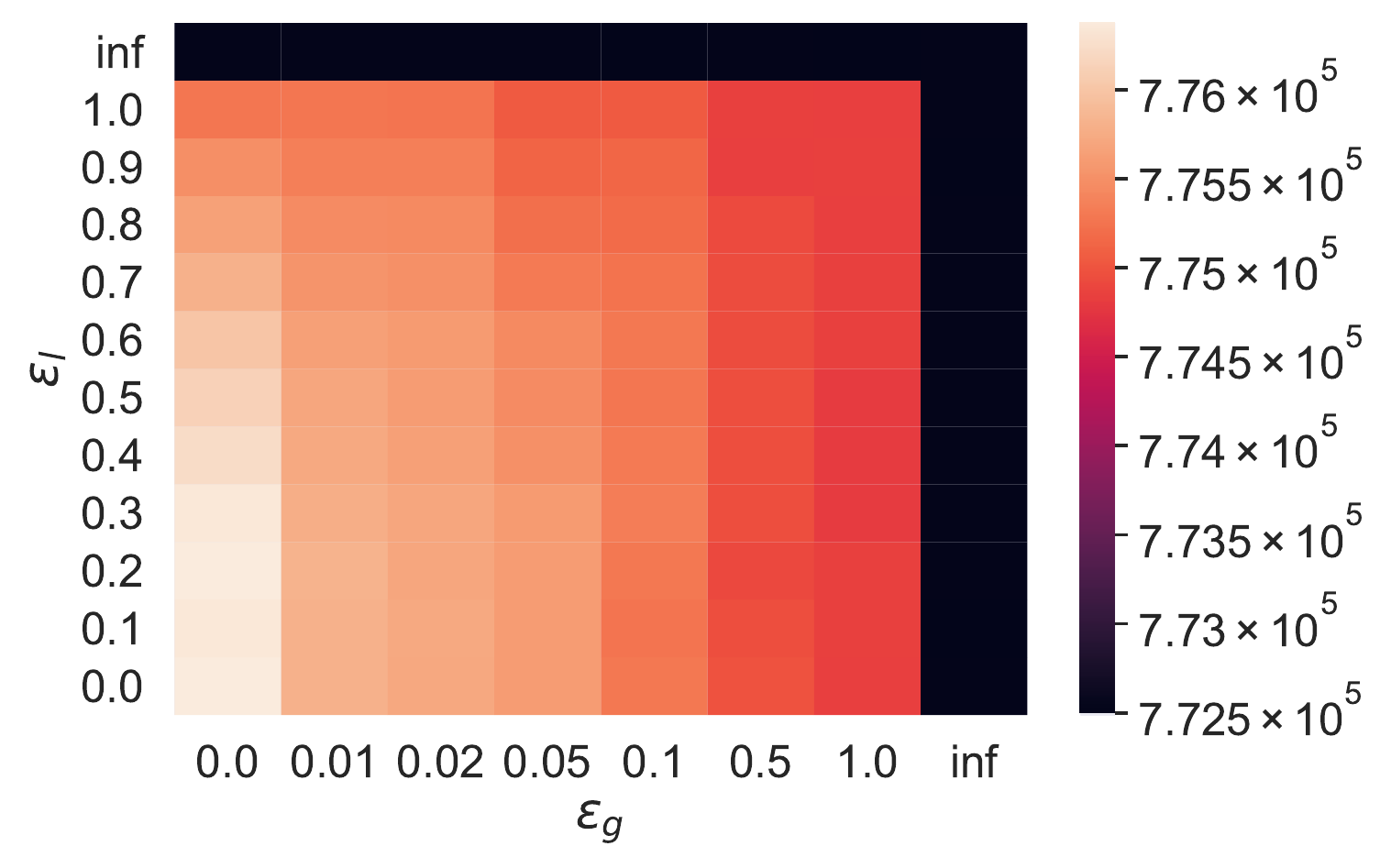}
        }
    \subfloat[{\tt Orkut} influence score\label{fig:orkut-score}]{%
        \includegraphics[width=0.33\linewidth]{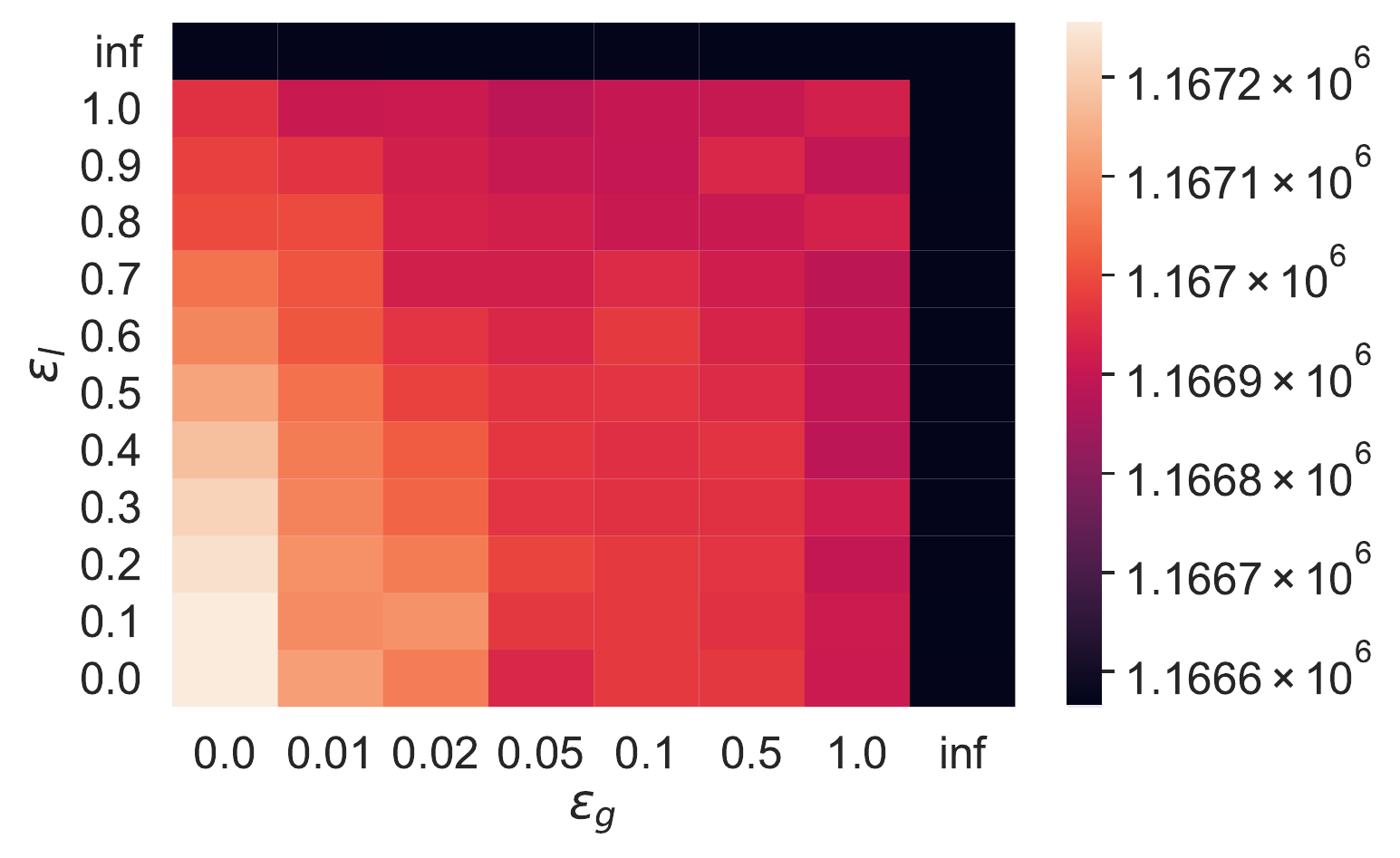}
        }
    \subfloat[{\tt Pokec} influence score\label{fig:pokec-score}]{%
        \includegraphics[width=0.33\linewidth]{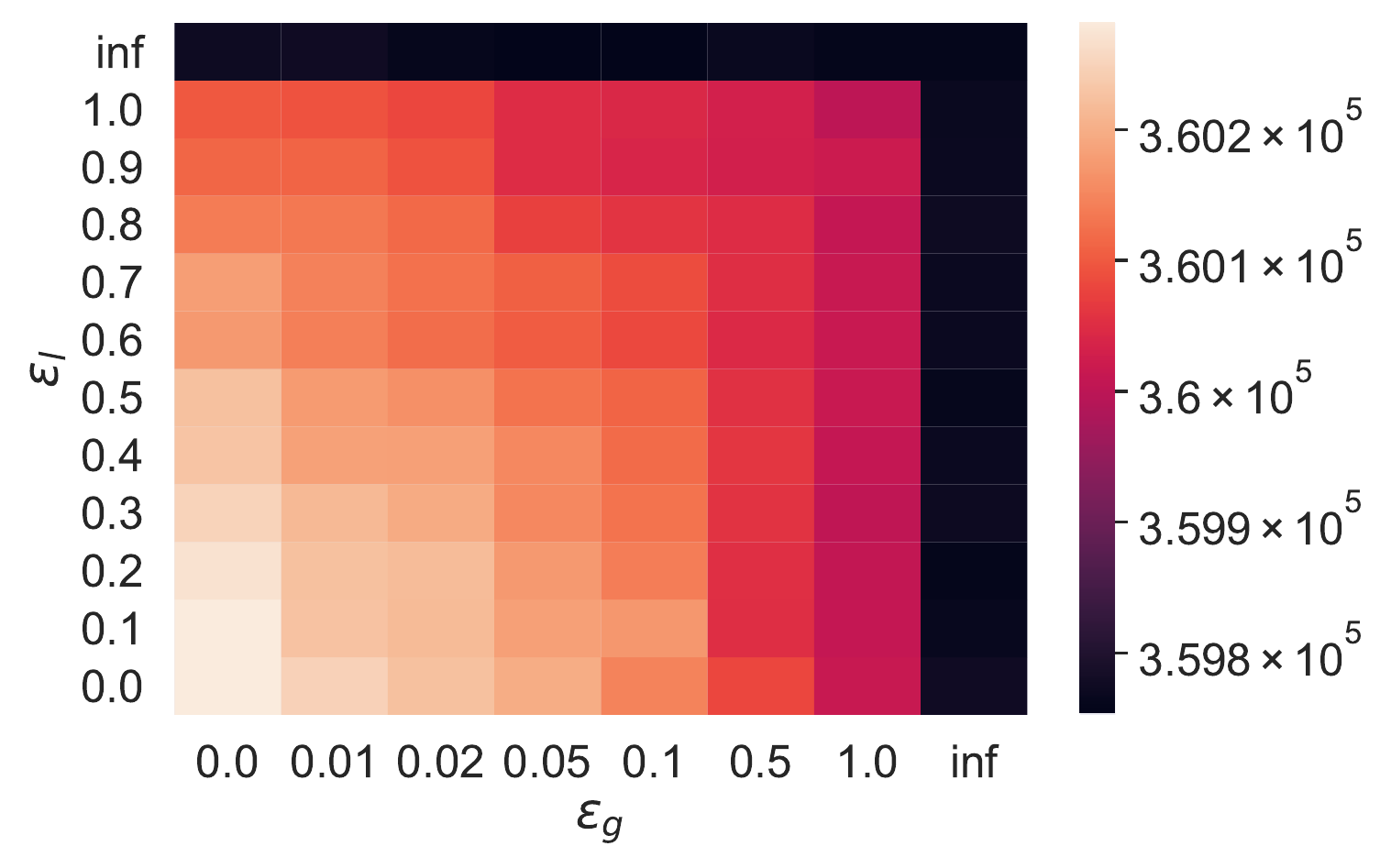}
        }\\
    \end{center}
    \caption{Effect of $\epsilon$ parameters on \acro(${\cal J} = 256$, $\epsilon_c=0.02$) performance, using $\tau=18$ threads. Lighter shades are better. }\label{fig:parameter-smallmultiples} 
\end{figure*}    

Figure~\ref{fig:sketch-saturation} shows the effect of register saturation by comparing two \acro variants; the first one rebuilds a new sketch to choose each seed vertex $s$, i.e., the {\bf else} part in lines~\ref{line:else1}--\ref{line:else2} of Algorithm~\ref{algo:main} is executed for every iteration of the {\bf for} loop at line~\ref{line:for}. This sketch is built on the residual graph $G \setminus R_G(S)$, which remains after the current seed set's reachability is removed. The second variant builds a  sketch only once at the beginning and employs it through the IM kernel, i.e., the {\bf else} part is never executed. The figure shows that the latter's seed selection quality is comparable to that of the former for the first few seed vertices. However, a significant reduction in the quality is observed for the later vertices. Furthermore, the former approach's quality is on par with the expensive Greedy algorithm's quality, which computes actual reachability sets. This shows that sketch-based estimation can perform as well as the accurate but expensive approach. Note that rebuilding also allows \acro to work on a smaller problem for the following seed vertex selection since we remove the already influenced vertices from sample subgraphs and work on the remaining subgraphs. 

Although its quality is on par with the traditional algorithm, the variant which rebuilds a sketch for all the seed vertex selections can be expensive. Here, we leverage an error-adaptive approach by rebuilding them when a significant cardinality estimation error is observed. 
The estimation error is calculated as follows; we store the influence score after each sketch rebuild in $\varsigma$~(line~\ref{line:else2} of Algorithm~\ref{algo:main}). Let $\sigma$ be the real influence for the seed set $S$ including the selected vertex. We first compute the marginal influence gain $\delta = \sigma - \varsigma$, which is the additional influence obtained since the last sketch rebuilt. Note that $e$, computed at line~\ref{line:e} is the sketch estimate for this value. \acro computes the local estimation error $err_l = |(e-\delta) / \delta|$ and the global error threshold $err_g=|(e-\delta) / \sigma|$. The sketches are assumed to be fresh if 
the local estimation error $err_l$ is smaller than a local threshold $\epsilon_{l}$ or the global error threshold $err_g=|(e-\delta) / \sigma|$ is smaller than a global threshold $\epsilon_{g}$.

 


The use of two different, local and global, thresholds allows the algorithm to rebuild the sketches after significant local errors and skip this expensive process if the estimation error is insignificant compared to the total influence. As explained above, when the rebuilding is skipped, \acro only updates $M_{S'}$ by merging it with the candidate vertex's sketch. Hence, the selected threshold values, $\epsilon_{l}$ and $\epsilon_{g}$, have a significant impact on the performance. Setting $\epsilon_{l} = \epsilon_{g} = 0$ means that the algorithm always rebuilds. Conversely, setting $\epsilon_l = \epsilon_{g} = \infty$ will make \acro fast since sketches are built only once. However, the influence scores will suffer, which is already shown by Fig.~\ref{fig:sketch-saturation}. To evaluate the interplay and find the thresholds that yield a nice tradeoff, we conducted a grid search in which \acro's execution time and influence quality are measured for different parameters. The results of this preliminary experiment are shown in Figure~\ref{fig:parameter-smallmultiples}.  We found that the parameters $\epsilon_{l}=0.3$ and $\epsilon_{g}=0.01$ perform well on many datasets, both in terms of speed and quality. 

\subsection{Implementation Details}


To efficiently process real-life graphs, \acro uses the Compressed Sparse Row~(CSR) graph data structure. In CSR, an array, $xadj$, holds the starting indices of the adjacency lists for each vertex, while another array, $adj$, holds the actual adjacency lists (i.e., the outgoing neighbors) one after another. Hence, the adjacency list of vertex $i$ is located in $adj$ at locations $adj[index[i], \ldots, index[i+1] - 1]$. 
    
The traditional two-step (sample-then-diffuse) computation model stores the (graph) data in a loosely coupled fashion.
While designing \acro, we fine-tuned it to be vectorization friendly, including its data layout and computation patterns. These design choices allow us to perform multiple operations, i.e., the same operations but on different data, at once. For instance, we keep all the memory registers of a single vertex from different simulations adjacent, and this allows the efficient use of vectorized computation hardware while performing lines~\ref{ln:vec1}--\ref{ln:update} of Algorithm~\ref{algo:diffusion-step}. Also, random number generation, fused sampling, and sketch merging are vectorizable operations when the data are stored in a coupled way as in \acro.

    
\section{Experimental Results}\label{sec:evaluation}
We performed the experiments on a server with an 18-core {\tt Intel Xeon Gold 6140}, running at 2.3Ghz, and 250GB memory. The Operating System on the server is {\tt Ubuntu 16.04 LTS} with 5.4.0-48 kernel. The algorithms are implemented using {\tt C++20}, and compiled with {\tt GCC 9.2.0} with {\tt "-Ofast"} and {\tt "-march=native"} optimization flags. Multi-thread parallelization was achieved with {\tt OpenMP} pragmas. {\tt AVX2} instructions are utilized by handcrafted code with vector intrinsics.

\begin{table}[!ht]
\caption{Properties of networks used in the experiments}\label{tab:NetProps}
\centering
\scalebox{0.95}{
\begin{tabular}{ll||r|r|r|r}
& & No. of           & No. of    & Avg.  &  Avg.  \\
&Dataset & Vertices          & Edges     &    Weight         &  Degree               \\
\hline
\multirow{6}{*}{\rotatebox[origin=c]{90}{Undirected}}& {\tt Amazon} & 262,113 & 1,234,878 & 1.00 & 4.71 \\
&{\tt DBLP} & 317,081 & 1,049,867 & 1.00 & 3.31 \\
&{\tt NetHEP} & 15,235 & 58,892 & 1.83 & 3.87 \\
&{\tt NetPhy} & 37,151 & 231,508 & 1.28 & 6.23 \\
&{\tt Orkut} & 3,072,441 & 117,185,083 & 1.00 & 38.14 \\
&{\tt Youtube} & 1,134,891 & 2,987,625 & 1.00 & 2.63 \\

\hline
\multirow{6}{*}{\rotatebox[origin=c]{90}{Directed}}&{\tt Epinions} & 75,880 & 508,838 & 1.00 & 6.71 \\
&{\tt LiveJournal} & 4,847,571 & 68,993,773 & 1.00 & 14.23 \\
&{\tt Pokec} & 1,632,803 & 30,622,564 & 1.00 & 18.75 \\
&{\tt Slashdot0811} & 77,360 & 905,468 & 1.00 & 11.70 \\
&{\tt Slashdot0902} & 82,168 & 948,464 & 1.00 & 11.54 \\
&{\tt Twitter} & 81,306 & 2,420,766 & 1.37 & 29.77 
\end{tabular}
}
\end{table}
\subsection{Experiment Settings}

We performed the experiments on twelve graphs~(six undirected, six directed). For comparability, graphs that have been frequently used within the Influence Maximization literature are selected. These graphs are {\tt Amazon} co-purchase network~\cite{snapnets}, {\tt DBLP} co-laboration network~\cite{snapnets}, {\tt Epinions} consumer review trust network, {\tt LiveJournal}~\cite{snapnets}, {\tt NetHEP} citation network~\cite{MixGreedy}, {\tt NetPhy} citation network~\cite{MixGreedy}, {\tt Orkut}~\cite{snapnets}, {\tt Pokec} Slovakian poker game site friend network~\cite{snapnets}, {\tt Slashdot} friend-foe networks~(08-11, 09-11)~\cite{snapnets}, {\tt Twitter} list co-occurence network~\cite{snapnets}, and {\tt Youtube} friendship network~\cite{snapnets}. The properties of these graphs are given in Table~\ref{tab:NetProps}. 

Three diffusion settings are simulated for a comprehensive experimental evaluation; for each network, we use 
\begin{enumerate}
    \item constant edge weights $w = 0.005$,
    \item constant edge weights $w = 0.01$~(as in~\cite{kempe2003maximizing} and~\cite{MixGreedy}),
    \item constant edge weights $w = 0.1$~(as in~\cite{kempe2003maximizing}),
\end{enumerate}

We selected $w=0.005$ as a benchmark-setting to challenge \acro. Due to its diffusion algorithm's nature, \acro traverses vertices even if they are blocked, which happens faster when the graph is sparser. Also, for each live vertex, \acro processes the sample edges for all simulations. The other two settings are selected to emulate the experiments of~\cite{kempe2003maximizing} and~\cite{MixGreedy}.

\subsection{Performance Metrics}

The algorithms are evaluated based on (1) execution time, (2) influence score, and (2) maximum memory used. For Influence Maximization, there is a trade-off among these performance metrics; in one extreme, it is trivial to select random vertices as the seed set. In another, one can compute the reachability sets of every possible seed set of size $K$ and choose the best one. 
In all our experiments, the execution times are the wall times reported by the programs. All the methods we benchmarked exclude the time spent on reading files and preprocessing. We only left out the time to spend on reading files for \acro. We allowed all methods to utilize all the CPU cores in all benchmarks, except {\sc Tim+}, a single-threaded algorithm. The memory use reported in this paper is the {\em maximum resident set sizes}~(RSS), which are measured using GNU {\tt time} command.

Since the algorithms may use different methods to measure the influence, the reported influence scores may not be suitable for comparison purposes with high precision. Due to this reason, we implemented an oracle with a straightforward, sample-then-diffuse algorithm without any optimization mentioned. For sampling, the random values are generated by the 32-bit Mersenne Twister pseudo-random generator {\tt mt19937} from {\tt C++} standard library. The same independent oracle obtains all influence scores in this paper.

\subsection{Algorithms evaluated in the experiments}

We evaluated our method against three other state-of-the-art influence maximization algorithms, {\sc Tim+}, {\sc Skim}, and {\sc Imm}. The first algorithm focuses on the influence score, whereas the second is a sketch-based algorithm that takes the execution time into account. The third one is an approximation algorithm with a parameter to control the influence quality.

\begin{itemize} [leftmargin=0.3cm]
\item The Two-phased Influence Maximization~({\sc Tim+}) runs in two phases: {\em Parameter Estimation} which estimates the maximum expected influence and a parameter $\theta$ and {\em Node Selection} which randomly samples $\theta$ reverse reachability sets from $G$ and then derives a size-$K$ vertex-set $S$ that covers a large number of these sets~\cite{tim}. The algorithm has a parameter $\epsilon$ which allows a trade-off between the seed set quality and execution time. In our experiments, we set  $\epsilon=0.3$ to have a high-quality influence maximization baseline. We also experimented with $\epsilon = 1.0$ as suggested, which gives around $7\times$ speedup on average but a reduction on the influence score up to $6\%$.

\item The Sketch-based Influence Maximization~({\sc Skim}) uses a combined bottom-$k$ min-hash reachability sketch~\cite{bottomk} to estimate the influence scores of the seed sets~\cite{cohen2014sketch}. As suggested by the authors, in this work, we employ {\sc Skim} with $k = 64$ and $\ell = 64$ sampled subgraphs. The implementation (from the authors) is partially parallelized and leverages multicore processors. 

\item Minutoli et al.'s {\sc Imm} is a high-performance, parallel algorithm that efficiently produces accurate seed sets~\cite{minutoli2019fast}. 
It is an approximation method that improves the Reverse Influence Sampling~(RIS)~\cite{borgs2014maximizing} algorithm by eliminating the need for the threshold to be used. 
We have used $\epsilon = 0.5$ as suggested in the original paper, where $\epsilon$ is a user-defined parameter to control the approximation boundaries.
\end{itemize}


\begin{table*} 
    \caption{\acro execution times~(in secs), influence scores, and memory use~(in GBs)  on the networks with $K = 50$ seeds using $\tau=18$ threads and constant edge weights $w=0.005$. Influence scores are given relative to \acro{}. The runs that did not finish due to high memory use shown as "-".
    }
    \label{tab:timings005}
    \centering
    \scalebox{0.94}{
\begin{tabular}{l|rrrr|rrrr|rrrr}
\toprule
{} & \multicolumn{4}{c|}{Time} & \multicolumn{4}{c|}{Influence Score} & \multicolumn{4}{c}{Memory} \\
Method & {\sc Hyper} &    {\sc Tim+} &    {\sc Imm} &  {\sc Skim} & {\sc Hyper} &    {\sc Tim+} &     {\sc Imm} &    {\sc Skim} & {\sc Hyper} &  {\sc Tim+} &  {\sc Imm} & {\sc Skim} \\

Dataset      & {\sc Fuser} &         &         &         & {\sc Fuser} &         &         &         & {\sc Fuser} &       &       &      \\

\midrule
{\tt Amazon}       &       1.30 &  124.38 &   5.58 & 63.73 &       96.9 &  1.041$\times$&  1.000$\times$&  0.562$\times$&       0.17 & 21.62 & 0.90 & 6.78 \\
{\tt DBLP}         &       1.61 &  177.99 &   5.71 & 28.24 &      106.6 &  1.068$\times$&  1.027$\times$&  1.036$\times$&       0.27 & 31.24 & 0.95 & 3.12 \\
{\tt Epinions}     &       1.11 &   12.16 &   0.50 &  8.29 &      635.3 &  1.026$\times$&  1.001$\times$&  0.939$\times$&       0.06 &  0.78 & 0.07 & 1.04 \\
{\tt LiveJournal}  &      13.25 & 4172.69 & 118.82 & 19.35 &    37174.1 &  1.010$\times$&  0.995$\times$&  0.957$\times$&       3.97 & 27.43 & 2.78 & 2.00 \\
{\tt NetHEP}       &       0.31 &    2.22 &   0.31 &  1.96 &       80.2 &  1.065$\times$&  0.993$\times$&  0.871$\times$&       0.01 &  0.80 & 0.04 & 0.33 \\
{\tt NetPhy}       &       0.39 &    7.40 &   0.40 &  1.00 &      124.5 &  1.042$\times$&  0.999$\times$&  0.777$\times$&       0.03 &  2.01 & 0.08 & 0.16 \\
{\tt Orkut}        &      30.22 &       - & 780.22 & 41.82 &   158842.6 &       - &  0.997$\times$&  1.001$\times$&       5.19 &     - & 7.81 & 1.77 \\
{\tt Pokec}        &      11.05 &  149.34 &   5.04 & 18.40 &     1095.1 &  1.032$\times$&  1.027$\times$&  0.925$\times$&       1.57 & 10.16 & 1.40 & 1.74 \\
{\tt Slashdot0811} &       1.18 &    6.93 &   0.40 &  1.00 &      576.4 &  1.015$\times$&  0.983$\times$&  0.942$\times$&       0.06 &  0.76 & 0.09 & 0.13 \\
{\tt Slashdot0902} &       1.14 &    6.96 &   0.40 &  1.17 &      610.5 &  1.022$\times$&  0.998$\times$&  0.953$\times$&       0.06 &  0.71 & 0.08 & 0.14 \\
{\tt Twitter}      &       1.10 &  171.17 &   5.40 &  1.60 &     3458.7 &  1.006$\times$&  0.990$\times$&  0.942$\times$&       0.09 &  2.02 & 0.12 & 0.15 \\
{\tt Youtube}      &       1.95 &   46.61 &   2.42 & 13.24 &     1820.8 &  1.025$\times$&  1.013$\times$&  1.000$\times$&       0.73 &  3.96 & 0.48 & 1.39 \\
\bottomrule
Norm. arit. mean &&69.39$\times$&4.37$\times$&8.13$\times$&&1.032$\times$&1.002$\times$&0.909$\times$& &42.60$\times$&2.05$\times$&9.76$\times$ \\ 
Norm. geo. mean &\multicolumn{1}{c}{all}&28.19$\times$&1.69$\times$&3.47$\times$&\multicolumn{1}{c}{all}&1.032$\times$&1.002$\times$&0.899$\times$&\multicolumn{1}{c}{all}&22.75$\times$&1.65$\times$&3.58$\times$ \\ 
Norm. max perf&\multicolumn{1}{c}{1.00$\times$}&314.92$\times$&25.82$\times$&49.02$\times$&\multicolumn{1}{c}{1.000$\times$}&1.068$\times$&1.027$\times$&1.036$\times$&\multicolumn{1}{c}{1.00$\times$}&127.18$\times$&5.29$\times$&39.88$\times$ \\ 
Norm. min perf&&5.88$\times$&0.34$\times$&0.85$\times$&&1.006$\times$&0.983$\times$&0.562$\times$&&5.43$\times$&0.66$\times$&0.34$\times$
\end{tabular}
    }
    \end{table*}
    
    \begin{table*} 
    \caption{\acro execution times~(in secs), influence scores, and memory use~(in GBs) on the networks with $K = 50$ seeds using $\tau=18$ threads and constant edge weights $w=0.01$. Influence scores are given relative to \acro{}. The runs that did not finish due to high memory use shown as "-".}
    \label{tab:timings01}
    \centering
    \scalebox{0.92}{

\begin{tabular}{l|rrrr|rrrr|rrrr}
\toprule
{} & \multicolumn{4}{c|}{Time} & \multicolumn{4}{c|}{Score} & \multicolumn{4}{c}{Memory} \\
Method & {\sc Hyper} &    {\sc Tim+} &    {\sc Imm} &  {\sc Skim} & {\sc Hyper} &    {\sc Tim+} &     {\sc Imm} &    {\sc Skim} & {\sc Hyper} &  {\sc Tim+} &  {\sc Imm} & {\sc Skim} \\

Dataset      & {\sc Fuser} &         &         &         & {\sc Fuser} &         &         &         & {\sc Fuser} &       &       &      \\
\midrule
{\tt Amazon }       &       0.96 &  107.94 &    3.28 &  59.32 &      152.7 &  1.037$\times$&  1.024$\times$&  0.390$\times$&       0.17 & 18.11 &  0.55 & 6.40 \\
{\tt DBLP }         &       0.73 &   73.37 &    2.85 &  18.71 &      233.5 &  1.043$\times$&  1.005$\times$&  0.997$\times$&       0.27 & 11.92 &  0.52 & 2.05 \\
{\tt Epinions }     &       0.82 &  112.08 &    3.78 &   5.07 &     2480.1 &  1.006$\times$&  0.983$\times$&  0.984$\times$&       0.06 &  1.95 &  0.10 & 0.68 \\
{\tt LiveJournal }  &      16.72 &       - &  386.37 &  16.23 &   155375.8 &       - &  0.996$\times$&  0.993$\times$&       3.97 &     - &  6.64 & 1.45 \\
{\tt NetHEP }       &       0.26 &    1.84 &    0.23 &   1.89 &      129.1 &  1.036$\times$&  0.997$\times$&  0.826$\times$&       0.01 &  0.60 &  0.03 & 0.31 \\
{\tt NetPhy }       &       0.24 &    3.33 &    0.23 &   0.86 &      320.5 &  1.010$\times$&  0.985$\times$&  0.732$\times$&       0.03 &  0.67 &  0.05 & 0.12 \\
{\tt Orkut }        &      42.37 &       - & 1870.35 & 114.82 &   650157.1 &       - &  1.000$\times$&  1.000$\times$&       5.19 &     - & 20.13 & 3.43 \\
{\tt Pokec }        &      11.65 & 4148.03 &   88.89 &   7.25 &    44685.8 &  1.004$\times$&  0.996$\times$&  0.988$\times$&       1.57 & 39.98 &  2.09 & 0.78 \\
{\tt Slashdot0811 } &       0.84 &  102.96 &    3.70 &   0.87 &     2882.1 &  1.003$\times$&  0.984$\times$&  0.976$\times$&       0.06 &  2.01 &  0.10 & 0.08 \\
{\tt Slashdot0902 } &       0.90 &  129.31 &    4.19 &   0.88 &     3061.5 &  1.008$\times$&  0.992$\times$&  0.980$\times$&       0.06 &  2.42 &  0.11 & 0.08 \\
{\tt Twitter }      &       0.90 &  390.36 &   10.96 &   1.22 &     9628.6 &  1.004$\times$&  0.992$\times$&  0.978$\times$&       0.09 &  4.91 &  0.23 & 0.09 \\
{\tt Youtube }      &       2.18 &  534.41 &   14.86 &  16.97 &     9042.7 &  1.009$\times$&  0.994$\times$&  1.006$\times$&       0.73 &  7.10 &  0.48 & 1.73 \\
\bottomrule
Norm. arit. mean &&167.17$\times$&9.73$\times$&9.99$\times$&&1.016$\times$&0.996$\times$&0.904$\times$& &42.91$\times$&2.09$\times$&8.26$\times$ \\ 
Norm. geo. mean &\multicolumn{1}{c}{all}&100.14$\times$&5.45$\times$&3.58$\times$&\multicolumn{1}{c}{all}&1.016$\times$&0.996$\times$&0.879$\times$&\multicolumn{1}{c}{all}&36.26$\times$&1.91$\times$&2.82$\times$ \\ 
Norm. max perf&\multicolumn{1}{c}{1.00$\times$}&433.73$\times$&44.14$\times$&61.79$\times$&\multicolumn{1}{c}{1.000$\times$}&1.043$\times$&1.024$\times$&1.006$\times$&\multicolumn{1}{c}{1.00$\times$}&106.53$\times$&3.88$\times$&37.65$\times$ \\ 
Norm. min perf&&7.08$\times$&0.89$\times$&0.62$\times$&&1.003$\times$&0.983$\times$&0.390$\times$&&9.73$\times$&0.66$\times$&0.37$\times$ 
\end{tabular}
    }
    \end{table*}
    \begin{table*} 
    \caption{\acro execution times~(in secs), influence scores, and memory use~(in GBs) on the networks with $K = 50$ seeds using $\tau=18$ threads and constant edge weights $w=0.1$. Influence scores are given relative to \acro{}. The runs that did not finish due to high memory use shown as "-".}
    \label{tab:timings1}
    \centering
    \scalebox{0.92}{

\begin{tabular}{l|rrrr|rrrr|rrrr}
\toprule
{} & \multicolumn{4}{c|}{Time} & \multicolumn{4}{c|}{Score} & \multicolumn{4}{c}{Memory} \\
Method & {\sc Hyper} &    {\sc Tim+} &    {\sc Imm} &  {\sc Skim} & {\sc Hyper} &    {\sc Tim+} &     {\sc Imm} &    {\sc Skim} & {\sc Hyper} &  {\sc Tim+} &  {\sc Imm} & {\sc Skim} \\

Dataset      & {\sc Fuser} &         &         &         & {\sc Fuser} &         &         &         & {\sc Fuser} &       &       &      \\
\midrule
{\tt Amazon}       &       0.69 &  133.22 &    1.98 &  23.62 &    11797.0 &  1.006$\times$&  0.990$\times$&  0.815$\times$&       0.17 &   5.49 &  0.23 & 2.59 \\
{\tt DBLP }         &       0.54 & 1368.83 &   14.54 &   6.30 &    48549.9 &  1.001$\times$&  0.995$\times$&  1.001$\times$&       0.27 &  35.19 &  1.06 & 0.65 \\
{\tt Epinions }     &       0.26 &  439.33 &    5.46 &   9.42 &    18409.9 &  1.000$\times$&  0.998$\times$&  0.997$\times$&       0.06 &  12.17 &  0.39 & 1.18 \\
{\tt LiveJournal }  &      10.10 &       - & 1071.30 &  65.73 &  2134726.0 &       - &  1.000$\times$&  1.000$\times$&       3.97 &      - & 65.49 & 1.40 \\
{\tt NetHEP }       &       0.10 &   14.18 &    0.33 &   0.61 &     2461.7 &  1.002$\times$&  0.975$\times$&  0.899$\times$&       0.01 &   1.02 &  0.04 & 0.10 \\
{\tt NetPhy }       &       0.16 &  107.52 &    1.53 &   0.34 &     8339.5 &  1.007$\times$&  0.994$\times$&  0.975$\times$&       0.03 &   3.84 &  0.13 & 0.03 \\
{\tt Orkut }        &      16.55 &       - & 1964.83 & 446.92 &  2692366.5 &       - &  1.000$\times$&  1.000$\times$&       5.19 &      - & 71.94 & 9.68 \\
{\tt Pokec }        &       4.90 &       - &  514.79 &  31.80 &  1034859.8 &       - &  1.000$\times$&  1.000$\times$&       1.57 &      - & 26.46 & 0.98 \\
{\tt Slashdot0811 } &       0.21 &  677.49 &    7.34 &   2.44 &    25871.8 &  1.000$\times$&  1.000$\times$&  0.999$\times$&       0.06 &  19.10 &  0.59 & 0.25 \\
{\tt Slashdot0902 } &       0.23 &  695.12 &    7.99 &   2.35 &    27519.5 &  1.000$\times$&  1.000$\times$&  0.999$\times$&       0.06 &  18.45 &  0.66 & 0.24 \\
{\tt Twitter }      &       0.33 & 1897.50 &   16.09 &   1.62 &    55327.3 &  1.000$\times$&  0.998$\times$&  0.998$\times$&       0.09 &  34.56 &  1.04 & 0.05 \\
{\tt Youtube }      &       1.12 & 7158.56 &   60.59 &  30.57 &   171392.9 &  1.000$\times$&  0.999$\times$&  1.001$\times$&       0.73 & 139.12 &  4.19 & 2.88 \\
\bottomrule
Norm. arit. mean &&2624.61$\times$&47.17$\times$&15.37$\times$&&1.002$\times$&0.996$\times$&0.974$\times$& &199.54$\times$&8.79$\times$&5.32$\times$ \\ 
Norm. geo. mean &\multicolumn{1}{c}{all}&1449.41$\times$&27.87$\times$&11.06$\times$&\multicolumn{1}{c}{all}&1.002$\times$&0.996$\times$&0.972$\times$&\multicolumn{1}{c}{all}&163.77$\times$&7.15$\times$&2.63$\times$ \\ 
Norm. max perf&\multicolumn{1}{c}{1.00$\times$}&6391.57$\times$&118.72$\times$&36.23$\times$&\multicolumn{1}{c}{1.000$\times$}&1.007$\times$&1.000$\times$&1.001$\times$&\multicolumn{1}{c}{1.00$\times$}&384.00$\times$&16.85$\times$&19.67$\times$ \\ 
Norm. min perf&&141.80$\times$&2.87$\times$&2.13$\times$&&1.000$\times$&0.975$\times$&0.815$\times$&&32.29$\times$&1.35$\times$&0.35$\times$ 
\end{tabular}

    }
    \end{table*}

\subsection{Comparing \acro with State-of-art}
To compare the run time, memory use, and quality of \acro with those of the state-of-the-art, 
we perform experiments using the following parameters controlling the quality of the seed sets: {\sc Tim+} ($\epsilon=0.3$), {\sc Imm} ($\epsilon=0.5$), {\sc Skim} ($l=64,k=64$). 
In fact, one of the drawbacks of \acro is that it does not have a direct control over the approximation factor, whereas {\sc Tim+} and {\sc Imm} have one. Still, \acro can control the quality indirectly by tuning the number of Monte-Carlo simulations ${\cal J}$ which also increases the number of sketches used per vertex. In the experiments, we set ${\cal J} = 256$. In addition, as explained in the previous section, we use a global error threshold $\epsilon_{g} = 0.01$, a local error threshold as $\epsilon_l=0.3$, and the early-exit ratio as $\epsilon_{c}=0.02$.\looseness=-1 

We present the results in Tables~\ref{tab:timings005},~\ref{tab:timings01} and~\ref{tab:timings1} for edge weights $w = 0.005$, $0.01$, and $0.1$, respectively. The top part of each table shows the  results for the networks, and the bottom four rows are the arithmetic mean, geometric mean, maximum and minimum, respectively, of the scores after they are normalized w.r.t. those of \acro's scores. In all tables, for the execution time (2--5) and memory (10--13) columns, lower values are better. For the influence scores, i.e., for columns 6--9, higher values are better. 

The tables show that for small and relatively sparser graphs such as {\tt NetHep}, {\tt NetPhy}, {\tt DBLP} and {\tt Amazon}, {\sc Tim+}, the high-quality baseline, has $0.1\%$--$7\%$ more influence score compared to the proposed approach. Except {\tt DBLP}, the other sketch-based algorithm, {\sc Skim} also performs bad on these graphs. For {\tt NetPhy}, {\tt NetHEP}, and {\tt Amazon}, its influence scores are $10\%$--$44\%$ worse that those of \acro{}.  For the rest of the graphs, {\sc Tim+} is only up to $3\%$, $0.9\%$ and $0.1\%$ better in terms of influence for the edge weights $w = 0.005$, $0.01$, and $0.1$, respectively,  while being $69\times$, $167\times$, and $2624\times$ slower on average over all the graphs. It is clear that \acro{}'s influence performance is getting closer to that of {\sc Tim+} when $w$ increases. Indeed, when $w$ is small, e.g., $0.005$, it may have a hard time while catching potential influence paths; the probability an edge being captured is $1-(1-0.005)^{256}=0.72$. Using $\epsilon = 0.3$, {\sc Tim+} does not suffer from sparsity, but as the tables show, {\sc Skim} can suffer more. With respect to the execution-time performance, \acro is superior to other methods; for instance, when $w = 0.01$, it is $167\times$, $10\times$, and $10\times$ faster on average compared to {\sc Tim+}, {\sc Imm}, and {\sc Skim}, respectively. Although {\sc Imm} and {\sc Skim} look similar for $w = 0.01$ in terms of relative average execution time performance, for large graphs, {\sc Skim} is faster than  {\sc Imm}. When $w = 0.01$, the maximum execution times for $4148$, $1870$, and $114$ seconds for {\sc Tim+}, {\sc Imm}, and {\sc Skim}, respectively. For the proposed approach and with the same $w$, the maximum time spent is only $42$ seconds. 

\acro{}'s memory consumption is less compared to those of others. Furthermore, it stays the same for all experimental settings with different $w$ values. This is partly due to fused sampling; the memory consumption is linearly dependent only on the number of vertices in $G$. Both sketch-registers and {\em visited} information are stored per vertex. Hence, \acro's memory consumption stays constant for any simulation parameters or any number of edges. That is given ${\cal J}$, \acro's memory use is predictable for any graph. On the other hand, the other methods' memory consumptions tend to increase with $w$, and their behaviours change with different parameters and graphs. 

Overall, the performance characteristics of the proposed algorithm are  different from its state-of-the-art competitors. \acro's performance is highly affected by $G$'s diameter. 
For instance, for {\tt Pokec} with $w = 0.01$, the average diameter of the samples is 43, which makes \acro to lose its edge against its fastest competitor. On the other hand, with $w = 0.1$, the average diameter is only around 17, and \acro is six times faster than its nearest competitor. Indeed, its execution time decreases as the samples and the influence graph $G$ get denser. On the other hand, the other methods tend to get slower under these changes.

 \begin{figure}[!ht] 
     \centering
     \includegraphics[width=0.93\linewidth]{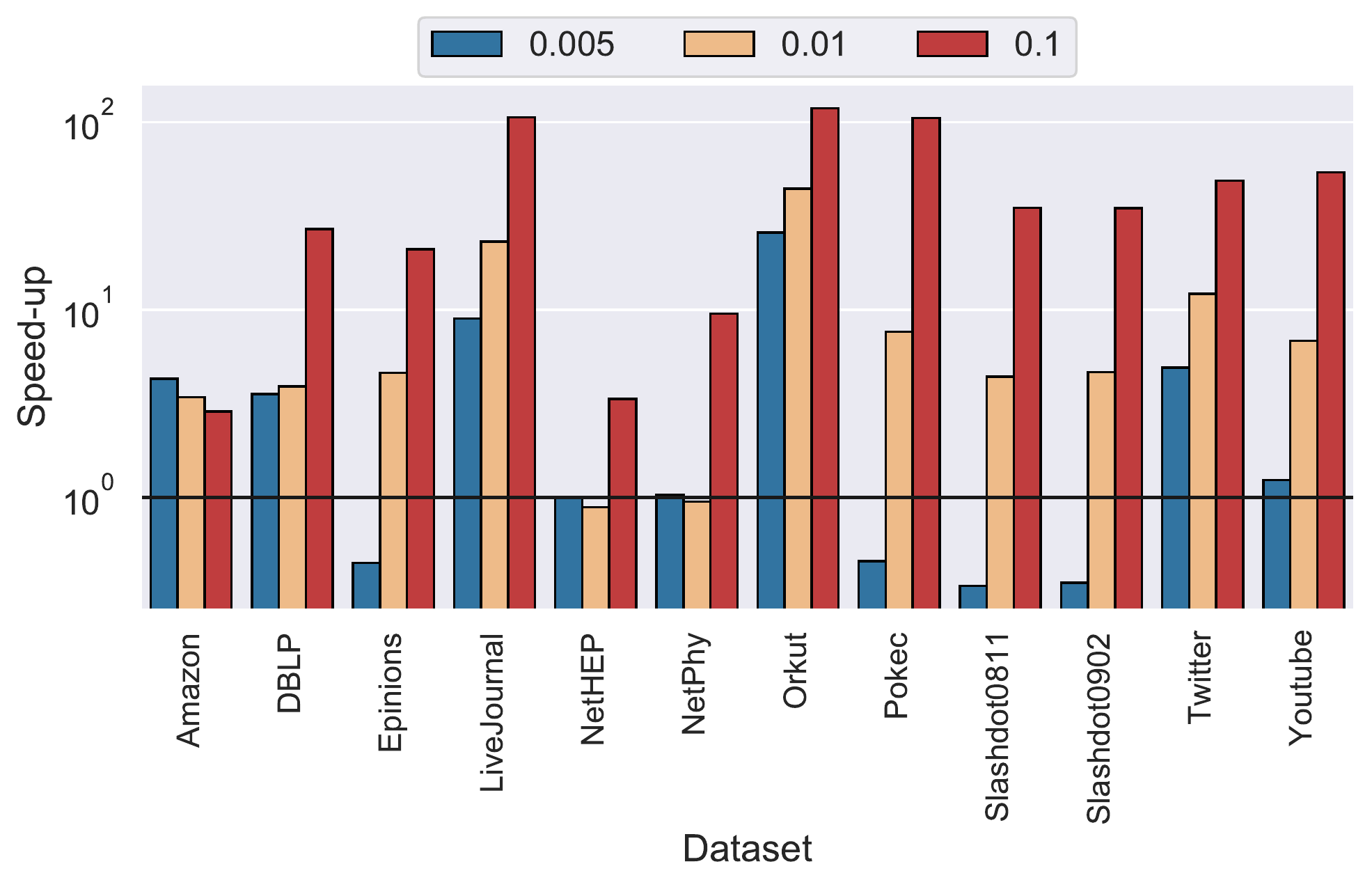}
   \centering \caption{Speedups obtained by \acro{}~(${\cal J} = 256$) over {\sc Imm}~($\epsilon\myeq 0.5$) using $\tau=18$ threads.
     \label{fig:vs-imm}} 
 \end{figure}

 \begin{figure}[!ht] 
     \centering
     \includegraphics[width=0.93\linewidth]{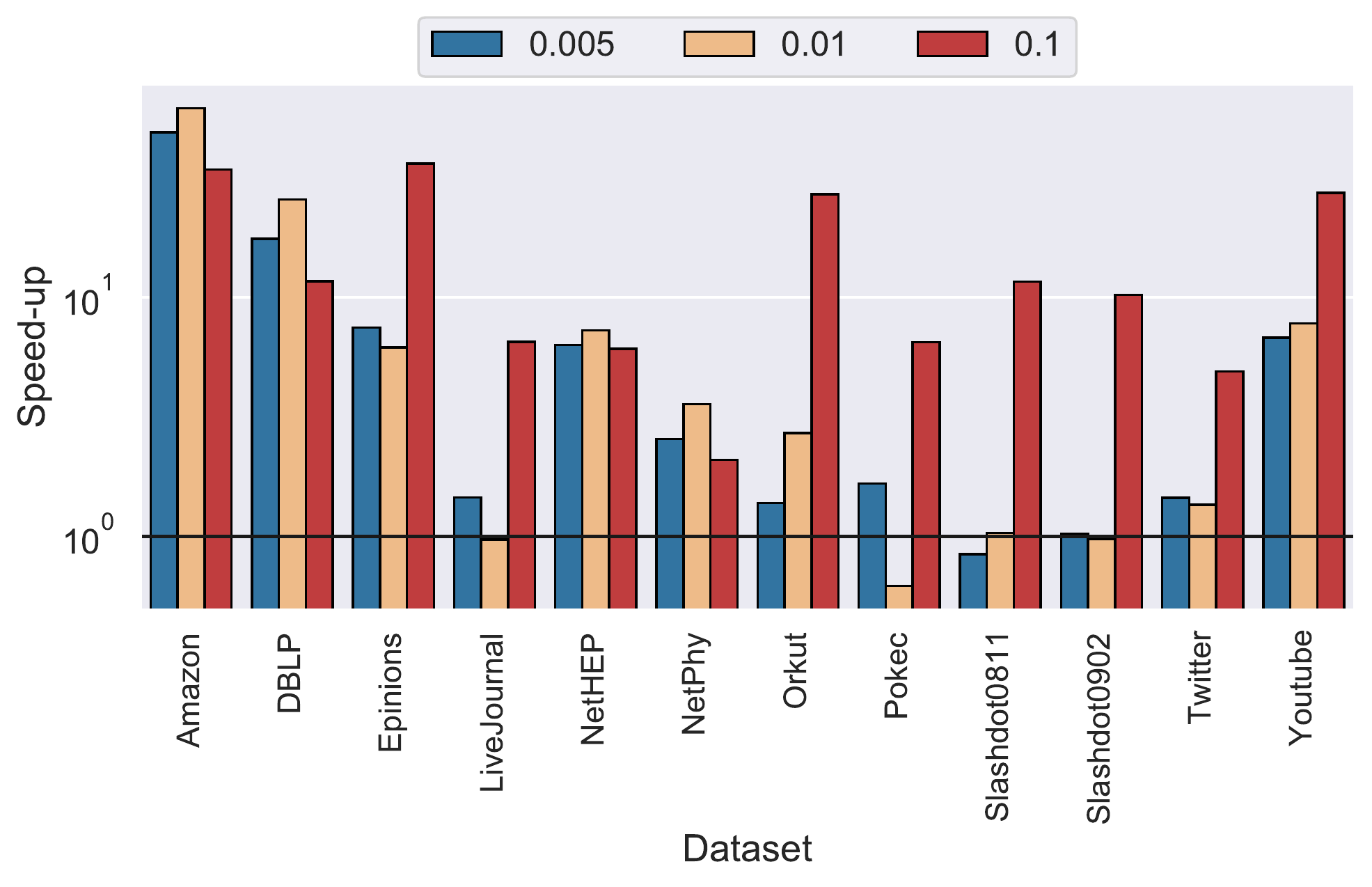}
   \centering \caption{Speedups obtained by \acro{}~(${\cal J} = 256$) over {\sc Skim}~($r\myeq 64,l\myeq 64$) using $\tau=18$ threads.
     \label{fig:vs-skim}} 
 \end{figure}
 
Figure~\ref{fig:vs-imm} shows the speedups of \acro over {\sc Imm} for all the graphs and all $w$ values. As described above, the relatively sparser setting $w=0.005$ is especially challenging due to the high diameter of the influence graph and low vector unit utilization. Even with this $w$ value, \acro{} is only slower by a few seconds and only when the influence is small. For larger graphs with larger influences \acro{} is much faster than {\sc Imm}. As explained before, for larger $w$, \acro{}'s execution-time performance is usually better, and its influence quality is on par with that of {\sc Imm}.


Figure~\ref{fig:vs-skim} compares \acro's execution-time performance with that of {\sc Skim}. As the figure shows, the proposed approach performs much better, both in terms of quality and speed in almost all settings. For the notorious {\tt Pokec} dataset, \acro performs better than {\sc Skim}, except for $w = 0.01$. The diameter of $G$ does not affect {\sc Skim}'s performance as much as \acro. {\sc Skim} is faster in this setting, but it has worse influence quality. In some settings such as {\tt Amazon} and $w = 0.01$, {\sc Skim} performs very poorly; only $39\%$ of the influence is achieved with respect to \acro. In addition, under the same setting, {\sc Skim} spends $59.3$ seconds whereas \acro finishes in less than one second.

\subsection{Scalability with multi-threaded parallelism}

\begin{figure*}[!ht] 
    \centering
    \includegraphics[width=0.93\linewidth]{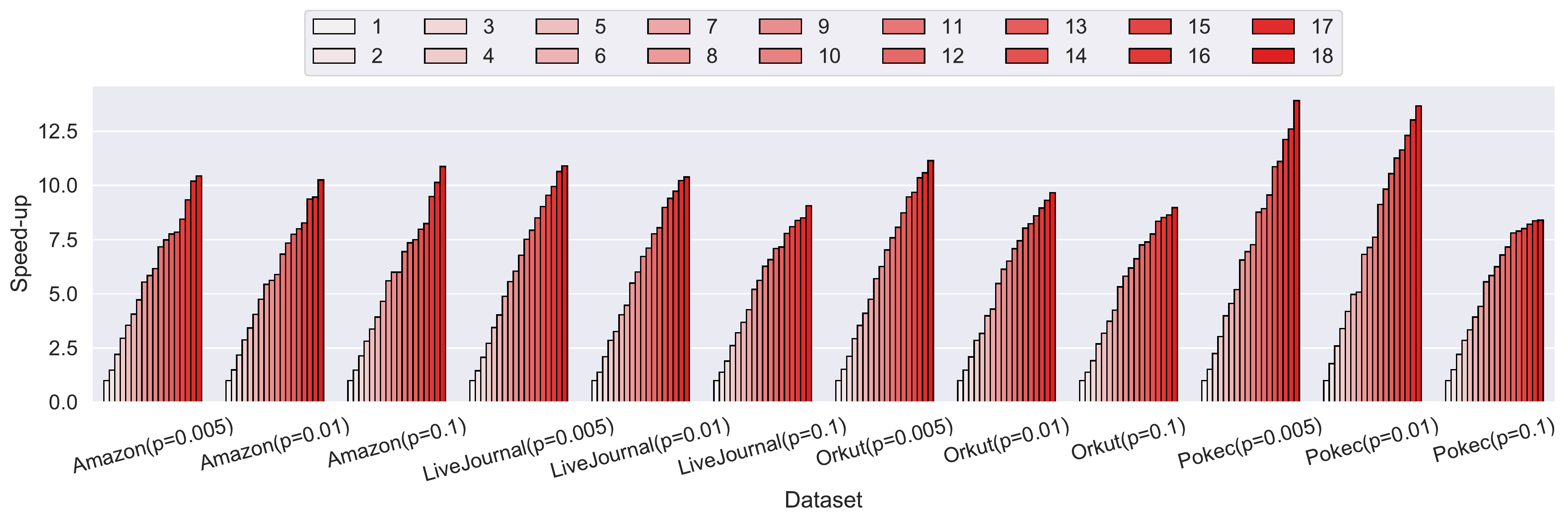}
   \centering \caption{Scaling of \acro with multiple threads on some of largest datasets in the benchmarks.
     \label{fig:scaling}} 
\end{figure*}

In our implementation, we used a {\em pull-based} approach in which the vertices~(processed at line~\ref{ln:inner_start} of Algorithm~\ref{algo:diffusion-step}) pull the influence, i.e., reachability set cardinality estimations, from their outgoing neighbors. A classical {\em push-based} approach, in which the vertices relay their influence to their outgoing neighbors could also be leveraged. However, the push-based approach makes a (target) vertex register potentially updated at the same time in different computation units. Specifically, the update operation~(corresponding to the one at line~\ref{ln:update} of Algorithm~\ref{algo:diffusion-step}) of the {\em pull-based} approach will be the cause of race conditions. One can easily argue that since we are already using sketches, and not computing exact cardinalities, such race conditions are acceptable and they will not reduce the quality of influence estimations. However, the performance may suffer due to false sharing. Figure~\ref{fig:scaling} shows \acro's speedup values obtained via a simple {\tt OpenMP} parallelization at line~\ref{ln:inner_start} of Algorithm~\ref{algo:diffusion-step}. Even though the pull-based diffusion shows a nice parallel performance, it is possible to implement \acro using other approaches such as the {\em queue-based} approach which may improve performance by only processing live vertices. The pull-based diffusion method is chosen due to its simplicity and scalability to many threads.

\section{Related Work}\label{sec:relatedwork}

Although they can be inferior in terms of influence, modern IM algorithms are shown to be quite fast compared to conventional simulation-based approaches such as {\sc MixGreedy}. Techniques such as using GPUs~\cite{IMGPU,curipples}, sketches for finding set intersections~\cite{cohen2014sketch,IPA}, reverse sampling to estimate the influence~\cite{borgs2014maximizing,minutoli2019fast}, and estimating the necessary number of simulations/samples required for each step~\cite{leskovec2009community} greatly reduces the execution times.
\acro borrows much from {\sc InFuseR}~\cite{infuser}, including hash-based fused sampling. {\sc InFuseR} computes influence by memoizing connected components for all vertices and only can work on {\em undirected} datasets. It also employs CELF optimization to reduce cardinality computations. On the other hand, \acro can process both directed and undirected graphs and uses the Flajolet–Martin sketches in a novel way to estimate cardinality and choose seed candidates. 

Sketch-based IM methods are cheaper compared to simulation-based methods. They usually pre-compute the sketches by processing the graph for evaluating the influence spread instead of running simulations repetitively. A popular method for sketch-based IM is {\sc Skim}~\cite{cohen2014sketch} by Cohen~et~al. {\sc Skim} uses combined bottom-$k$ min-hash reachability sketches~\cite{bottomk,cohen2015all}, built on $\ell$ sampled subgraphs, to estimate the influence scores of the seed sets. It is parallel in the sense that it uses {\tt OpenMP} parallelization during sketch utilization. However, the sketch building step is single-threaded. In this work, we choose Flajolet-Martin sketches~\cite{flajolet1985probabilistic} for their simplicity, suitability for vectorization and fused sampling, and hence, execution-time performance. 
 {\sc Skim} treats vertex/sample pairs as distinct elements and reduces edge traversals via their smallest ranks in bottom-$k$ ketches. On the other hand, \acro sees vertices as shared elements among the samples, and builds a sketch for each instance. This allows independent parallel processing of vertices and samples, fused sampling, and a better memory layout than the former. In addition, \acro does not require removing the reachable vertices from the samples. Instead, it uses a rebuilding strategy to improve the result quality.
 
The Independent Path Algorithm~(IPA)~\cite{IPA} by Kim~et~al. runs a proxy model and prunes paths with probabilities smaller than a given threshold in parallel. The approach only keeps a dense but small part of the network and scalable to only sparse networks. Liu~et~al. proposed IMGPU~\cite{IMGPU}, an IM estimation method by utilizing a bottom-up traversal algorithm. It performs a single Monte-Carlo simulation on many GPU threads to find the reachability of the seed set. It is $5.1\times$ faster than {\sc MixGreedy} on a CPU. The GPU implementation is up to $60\times$ faster with an average speedup of  $24.8\times$. 

Borgs~et~al.~\cite{borgs2014maximizing} proposed Reverse Influence Sampling~(RIS) which samples a fraction of all random reverse reachable sets. The number of necessary samples to find the seed set is calculated based on the number of visited vertices. The algorithm has an approximation guarantee of $(1-1/e-\epsilon)$. Minutoli et al. improved RIS and proposed {\sc Imm} that works on multi-threaded and distributed architectures~\cite{minutoli2019fast} with high efficiency. Recently, the authors extended the algorithm to work in a multi-GPU setting~\cite{curipples}.

Two-phased Influence Maximization~({\sc Tim+}) borrows ideas from RIS but overcomes its limitations with a novel algorithm~\cite{tim}. Its first phase computes a lower bound of the maximum expected influence over all size-$K$ node sets. It then uses this bound to derive a parameter $\theta$. In the second phase, it samples $\theta$ random RR sets from $G$, and then derives a size-$K$ node-set that covers a large number of RR sets.


Kumar and Calders~\cite{kumar2017information} proposed the Time Constrained Information Cascade Model and a kernel that works on the model using versioned HyperLogLog sketches. The algorithm computes the influence for all vertices in $G$ while performing a single pass over the data. The sketches are used for each time window to estimate active edges. On the other hand, \acro uses ${\cal J}$ sketches for each vertex to estimate the marginal influence and employs a rebuilding strategy for fast processing. \acro also utilizes fused sampling and error-adaptive rebuilding of sketches.

\section{Conclusion and Future Work}\label{sec:conclusion}

In this work, we propose a sketch-based Influence Maximization algorithm that employs fused sampling and error-adaptive rebuilding. We provide a fast implementation of the algorithm that utilizes multi-threading to exploit multiple cores. Also, we present a performance comparison with state-of-the-art IM algorithms on real-world datasets and show that \acro{} can be an order of magnitude faster while obtaining the same influence. 
In the future, we will extend our work to a distributed GPGPU setting to process graphs with billions of vertices and edges under a minute. 


\ifCLASSOPTIONcaptionsoff
  \newpage
\fi

\bibliographystyle{IEEEtran}
\bibliography{refs}
\begin{IEEEbiography}[{\includegraphics[width=1in,height=1.25in,clip,keepaspectratio]{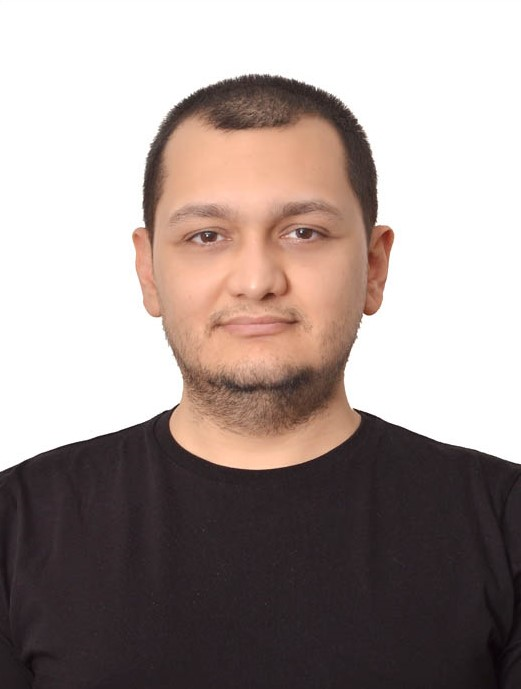}}]{G\"{o}khan~G\"{o}kt\"{u}rk} is a PhD candidate at the Faculty of Engineering and Natural Sciences in Sabancı University. He has received his BS and MS degrees from Sabancı University as well. He is interested in High Performance Computing, Parallel Programming, and Graph Processing.
\end{IEEEbiography}
\begin{IEEEbiography}[{\includegraphics[width=1in,height=1.25in,clip,keepaspectratio]{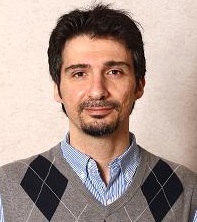}}]{Kamer Kaya} is an Asst. Professor at Sabancı University. After receiving his PhD from Bilkent University, he worked at CERFACS, France, as a post-graduate researcher in the Parallel Algorithms Project. He then joined the Ohio State University in September 2011 as a postdoctoral researcher, and in December 2013, he became a Research Assistant Professor in the Dept. of Biomedical Informatics.
His current research interests include Parallel Programming, High-Performance Computing, and Cryptography. 
\end{IEEEbiography}

\end{document}